%% file: main.tex
\documentclass[twocolumn,epjc3]{svjour3}       
\usepackage{graphicx,color}
\usepackage{amsmath,autobreak}
\usepackage{array}
\usepackage[utf8]{inputenc}
\usepackage[backend=bibtex,sorting=none]{biblatex}
\usepackage[sorting=none]{biblatex}
\usepackage[table]{xcolor}
\RequirePackage{ifpdf} 
\usepackage{float}
\usepackage{mathtools}
\usepackage{cases}
\usepackage{bm,amssymb}
\usepackage{amsbsy}
\usepackage{footnote}
\usepackage{bm}
\usepackage{pstricks}
\usepackage[final]{pdfpages} 

\usepackage{color}
\definecolor{urlblue}{rgb}{0.2,0.4,0.7}
\definecolor{citegreen}{rgb}{0,0.6,0.2}
\definecolor{linkred}{rgb}{0.9,0.2,0.1}
\usepackage{graphics}
\usepackage{etoolbox}
\usepackage{fixmath}
\usepackage{psfrag}
\usepackage{bm}
\usepackage{notoccite}
\usepackage{amsfonts}
\usepackage{autobreak}
\usepackage{marginnote}
\usepackage{widetext}
\usepackage{tabularx}
\usepackage{multirow, booktabs}
\usepackage{slashed}
\usepackage{array}

\RequirePackage[T1]{fontenc}
\smartqed 
\RequirePackage{graphicx}
\RequirePackage{mathptmx}
\RequirePackage{flushend}
\RequirePackage[colorlinks,citecolor=blue,urlcolor=blue,linkcolor=blue]{hyperref}

\usepackage{marginnote}
\usepackage{widetext}
\usepackage{scalefnt}
\usepackage{booktabs,multirow,tabularx}
\usepackage{rotating}
\usepackage{microtype}
\usepackage{makecell}
\usepackage{cancel}
\usepackage{mathtools}
\input{definitions.tex}

\usepackage{multirow, caption, booktabs}
\usepackage{slashed}
\usepackage{array}

\newcommand{\wh}[1]{\widehat{#1}}

\newcommand{\dd}{\mathop{}\!\mathrm{d}}

\newcommand{\dipole}[1]{\left. \mathcal{{J}}^{(2)}_{q\bar{q}}\right|_{N_f}}

\DeclareUnicodeCharacter{2212}{-}

\allowdisplaybreaks

\journalname{Eur. Phys. J. C}

\begin{document}

\title{\Large \bf Time-Like Heavy-Flavour Thresholds for Fragmentation Functions: the Light-Quark Matching Condition at NNLO}

\author{Christian Biello\thanksref{e1,addr1} 
      \and
         Leonardo Bonino\thanksref{e2,add2} 
 }
\thankstext{e1}{\href{mailto:biello@mpp.mpg.de}{biello@mpp.mpg.de}}
\thankstext{e2}{\href{mailto:leonardo.bonino@physik.uzh.ch}{leonardo.bonino@physik.uzh.ch}}

\institute{Max-Planck-Institut f\"ur Physik, Boltzmannstraße 8, 85748 Garching, Germany\label{addr1}  \and
Physik-Institut, Universit\"at Z\"urich, Winterthurerstrasse 190, CH-8057 Z\"urich, Switzerland \label{add2}
}
\date{}
\maketitle

\begin{abstract}
 Matching conditions are universal ingredients that describe how fragmentation functions change when heavy-flavour thresholds are crossed during the factorisation scale evolution. They are the last missing piece for a consistent description of observables with identified final-state hadrons at next-to-next-to leading order accuracy in quantum chromodynamics. We present an analytical form of the matching condition for light-flavour to hadron fragmentation function at next-to-next-to leading order. The derivation is performed by extending the formalism employed in the extraction of the next-to leading order matching conditions to the subsequent order, making use of $e^+e^-$ annihilation cross sections.
 We obtain the first non-trivial heavy-quark effect in the light-quark fragmentation functions and provide results in Mellin space.
\end{abstract}

\section{Introduction}
\label{sec:intro}
The production of identified hadrons in high energy collisions is described in quantum chromodynamics (QCD) through fragmentation functions (FFs). They parameterise the non-perturbative fragmentation of a parton (quark or gluon) into the identified hadron, which carries a fraction of the parent parton momentum \cite{Field:1976ve,Field:1977fa}. Thanks to the factorisation theorem, a differential cross section for the production of an identified hadron can be written as the convolution of a process-dependent coefficient function, encoding the hard scattering of the process, a process-independent fragmentation function, and up to two process-independent parton distribution functions (PDFs), according to the collider under consideration. FFs and PDFs obey Dokshitzer-Gribov-Lipatov-Altarelli-Parisi (DGLAP) evolution equations in their resolution scale, bridging the gap between the hard scattering scale and non-perturbative scales \cite{Altarelli:1977zs}.

Precision studies of processes with identified hadrons at colliders have received growing interest in the last years.  Important progress was made with regard to light hadron production (e.g. pions) thanks to the efforts in global FFs fits with recent results including $e^+e^-$ and approximate semi-inclusive deep-inelastic scattering (SIDIS) data \cite{Borsa:2022vvp,AbdulKhalek:2022laj}. The precision of these studies was in part limited due to the absence of SIDIS coefficient functions at next-to-next leading order (NNLO), which have only recently been computed \cite{PhysRevLett.132.251901,PhysRevLett.132.251902}.

Regarding heavy hadrons, such as $B$ and $D$ mesons, the state-of-the-art of FFs is less developed. Using the perturbative fragmentation function formalism, fits have been performed at NNLO to $e^+e^-$ data for $B$ \cite{Czakon:2022pyz,Bonino:2023icn} and $D$ mesons \cite{Bonino:2023icn} only recently. From a theoretical perspective, the calculation is very similar to the massless one, the only difference being that perturbative fragmentation functions are used. These encode logarithmically enhanced massive effects \cite{Mele:1990yq}, which are known up to NNLO \cite{Melnikov:2004bm,Mitov:2004du}. In order to properly account for soft gluon effects, the initial conditions must be resummed up to next-to-next-to leading logarithmic level \cite{Czakon:2022pyz,Aglietti:2006yf,Ridolfi:2019bch,Maltoni:2022bpy}. Fits to heavy hadrons without the inclusion of perturbative initial conditions have been performed up to NNLO in e.g.~\cite{Kneesch:2007ey,Salajegheh:2019nea}.
In both cases, the resummation of logarithmically enhanced collinear divergences is achieved by DGLAP evolution, in which heavy-flavour thresholds are crossed. Therefore a time-like variable flavour number scheme (VFNS) up to NNLO is required. At NLO the ingredients were computed in \cite{Cacciari_2005} and used to study heavy hadron FFs at NLO \cite{Cacciari_2006}. These matching conditions are currently implemented in Mellin space in the public evolution library \texttt{MELA} \cite{Bertone:2015cwa}. In addition to this matching, including heavy-quark threshold effects in the $e^+e^-$ coefficient functions reduces the discrepancy between theory and experimental data in the charm ratio \cite{Cacciari:2024kaa}. At NNLO, these matching conditions are still missing, yet crucial for achieving full NNLO accuracy on for example $D$-mesons FFs. This is also relevant when it comes to LHC phenomenology, as highlighted in studies involving $D$-meson production in association with a vector boson \cite{Caletti:2024xaw}.

From a broader perspective, these matching conditions are conceptually equivalent to their space-like counterparts, where crossing thresholds in PDFs evolution has had significant impact on global fits. The matching equations for PDFs \cite{Collins:1986mp} have long been known at NNLO level \cite{Buza:1996wv}, using the formalism of Operator Matrix Elements (OME) \footnote{The space-like results from the OME calculations cannot be directly mapped to their time-like counterparts. Indeed, these heavy-quark coefficient functions are expressed using dispersion integrals for off-shell forward Compton scattering, following the standard approach of the Operator Product Expansion (OPE) in deep inelastic scattering (DIS)~\cite{vanneerven1997c}. Therefore the time-like matching conditions are extracted from a physical process.}. Recently, all the necessary contributions for the next-to-NNLO ($\text{N}^3\text{LO}$) accuracy were computed \cite{Bierenbaum:2009zt, Bierenbaum:2009mv, Ablinger:2014vwa, Ablinger:2022wbb,Ablinger:2023ahe,Ablinger:2024xtt}. In addition, matching equations played a crucial role in providing evidence for intrinsic charm in the proton \cite{NNPDF:2023tyk,Ball:2022qks}. It is thus important to increase the level of accuracy when crossing flavour thresholds, in the time-like case too. Matching conditions are an ingredient of the time-like DGLAP evolution in the VFNS that can be implemented in fits to FFs independently on the choice of the initial conditions. 

The aim of this work is to pave the way for the calculation of these missing matching coefficients and provide first results.
Some of the ingredients needed for these calculations have already been computed in the context of the antenna subtraction formalism \cite{Gehrmann-DeRidder:2005btv,Gehrmann:2022pzd,Bonino:2024adk}, which can now accommodate fragmentation in hadronic collisions. We make use of some of these results and analytically compute the missing massive pieces. All massive contributions are given in small-mass limit, neglecting power corrections. In section \ref{sec:NLO} we review the NLO matching conditions focusing on details relevant to the NNLO extension. In section \ref{sec:masterformulae} we present the master formulae for all matching conditions.  In section \ref{sec:xsections} we provide results for the light quark matching condition, which employs existing massless antennae as well as novel massive corrections. Conclusions and outlooks are given in section \ref{sec:conclusions}. Details of the calculation of the massless and massive elementary cross-sections are reported in the appendices.

\section{Revising the NLO derivation}
\label{sec:NLO}

In this section we summarise the derivation of the threshold conditions at NLO accuracy closely following Ref. \cite{Cacciari_2005}. We consider the semi-inclusive production of an hadron $H$ in $e^+e^-$ annihilation
\begin{equation}
    e^+e^- \rightarrow\gamma^*\rightarrow H+ X,
\end{equation}
at a center-of-mass energy $Q$ much greater than the heavy-flavour mass $m$, where power-suppressed terms of order $m/Q$ can be neglected. On the other hand, $Q$ should not be arbitrarily large, ensuring that $\alpha_s \log\frac{Q}{m}$ remains  smaller than 1.
Due to the factorisation theorem, the FFs factor out in the differential cross section, at the cost of introducing an unphysical factorisation scale $\mu$.  \\

For scales $\mu$ much below the heavy-flavour threshold, the differential cross section for single hadron production can be written as a convolution of a parton-level cross-sections and FFs $D_j^{(n_L)}$, encoding the transition of the massless partons, namely the gluon and the quarks of $n_L$ flavours, into hadrons
\begin{align}
    j\in \mathbb{I}_{n_L}=\left\{ q_1, \bar{q}_1, \dots,  q_{n_L},\bar{q}_{n_L}, g\right\}.
\end{align}
The evolution of $D_j^{(n_L)}$ in terms of the scale $\mu$ is given by the DGLAP equation with $n_L$ flavours. Heavy-flavour effects are instead confined in the parton-level cross-section with a full mass dependence. This setup is called decoupling scheme, since the ultraviolet (UV) renormalisation is performed in the $\overline{\text{MS}}$ scheme for $n_L$ flavours and the divergences from heavy-quark loop are subtracted at zero-momentum. In contrast, for scales $\mu$ significantly larger than the mass $m$, the FFs of the heavy-quark $h$ and its antiparticle are introduced. Specifically we consider the set of functions $D_k^{(n)}$ for $k \in \mathbb{I}_n=\mathbb{I}_{n_L}\cup \{h,\bar h\}$ which obey DGLAP evolution equations for $n=n_L+1$ flavours. The renormalisation is perfomed in the full $\overline{\text{MS}}$ scheme for all flavours.

In the decoupling scheme, the differential cross section at $\mathcal{O}(\alpha_s)$ for hadron $H$ production with energy fraction $x=2E_H/Q$ is
\begin{align}
	\frac{\dd \sigma_H}{\dd x}=\int_x^1 \frac{\dd z}{z} \Bigg\{ \sum_{j\in \mathbb{I}_{n_L}} D_j^{(n_L)}\left(\frac{x}{z},\mu\right)\frac{\dd \sigma_j^{}(z,\mu)}{\dd z} \nonumber \\
     +D_g^{(n_L)}\left(\frac{x}{z},\mu\right)\frac{\dd \sigma_{h\bar{h}g^{\text{id.}}}(z,m)}{\dd z} \Bigg\}, \label{decNLO}
\end{align}
where $\dd \sigma_j$ is the cross-section for an identified parton $j$ with in principle heavy-flavour mass effects and $\dd \sigma_{h\bar{h}g^{\text{id.}}}$ describes the production of a massive heavy-quark pair in association with an identified gluon. From the massive cross-sections we retain only the leading contribution, namely the logarithmically enhanced and finite pieces, neglecting all power corrections in $m$.

Up to $\mathcal{O}(\alpha_s)$, in the first contribution of eq.~\eqref{decNLO} one can replace $\dd \sigma_j^{}$ with the cross-section for an identified parton $j$ when treating the heavy-quark $h$ as massless (i.e.~nominally heavy), since there are no fermion loop effects at NLO. This is denoted as $\dd \hat\sigma_{j}^{}$, where the hat stands for the massless treatment of the heavy-flavour. The last contribution in eq.~\eqref{decNLO} represents the cross-section for the process
\begin{equation}
	\gamma^*\left(q^{(Q)}\right)\rightarrow h\left(k_1^{(m)}\right)+\bar{h}\left(k_2^{(m)}\right)+g\left(k_p^{(0)}\right) \, , \label{eq:process}
\end{equation}
where we indicate the four-momenta of the state in parenthesis with its invariant mass as superscript. 

In the full (massless) $\overline{\text{MS}}$ scheme, the differential cross-section of eq.~\eqref{decNLO} becomes
\begin{align}
	\frac{\dd \sigma_H}{\dd x}=\int_x^1 \frac{\dd z}{z} \left\{ \sum_{k\in \mathbb{I}_{n}} D_k^{(n)}(\tfrac{x}{z},\mu)\frac{\dd \hat \sigma_k^{}(z,\mu)}{\dd z}\right\}, \label{MSNLO}
\end{align}
where all flavours, including the (nominally) heavy one, are treated as light. We stress that the quark $h$ is a massless particle in $\dd \hat \sigma_k$.

The matching conditions can be derived by taking the difference at NLO accuracy between the predictions in the two schemes \cite{Cacciari_2005}, namely eq.~\eqref{decNLO} and eq.~\eqref{MSNLO}. The contributions in the difference are collected in terms of the electromagnetic coupling constants of the quarks and one can ask for the vanishing behaviour of each independent coefficient. Considering the coefficients proportional to the light-quark charge ($Q_i^2$), one can easily infer that the difference between the light quark fragmentation functions is beyond NLO accuracy
\begin{equation}
	D_i^{(n)}(x,\mu)=D_i^{(n_L)}(x,\mu)+\mathcal{O}(\alpha_s^2), \hspace{0.5cm}i\in \mathbb{I}_{n_L}\smallsetminus\{g\} \, .
\end{equation}
In sections \ref{sec:masterformulae} and \ref{sec:xsections} we compute the first non-trivial correction to the previous equation.\\

The contribution that comes with the gluon fragmentation function instead is proportional to
\begin{align}
	& D_g^{(n)}\left(\frac{x}{z},\mu\right) \frac{\dd \hat \sigma_g^{(n)}(z,\mu)}{\dd z}-D_g^{(n_L)}\left(\frac{x}{z},\mu\right) \frac{\dd \hat \sigma_g^{(n_L)}(z,\mu)}{\dd z} \nonumber \\
 & = D_g^{(n)}\left(\frac{x}{z},\mu\right) \frac{\dd \hat \sigma_{h\bar{h}g^{\text{id.}}}(z,\mu)}{\dd z} \, ,
\end{align}
where we can set $D_g^{(n)}=D_g^{(n_L)}$ in the previous equation at NLO accuracy, defining the difference between the two differential cross-sections as $\dd \hat \sigma_{h\bar{h}g^{\text{id.}}}$. The latter is the probability of producing the massless quark pair of (nominally) heavy-flavour $h$ in association with an identified gluon. 

The leftover from the difference between the results in the two schemes is proportional to the heavy-quark charge ($Q_h^2$) and reads
\begin{align}
	&\int_x^1 \frac{\dd z}{z} D_g^{(n_L)}\left(\frac{x}{z},\mu\right) \left[ \frac{\dd \sigma_{h\bar{h}g^{\text{id.}}}(z,m)}{\dd z} - \frac{\dd \hat \sigma_{h\bar{h}g^{\text{id.}}}(z,\mu)}{\dd z} \right] \nonumber \\
	&\hspace{1cm}+\int_x^1 \frac{\dd z}{z} \sum_{k\in\{h,\bar{h}\}} D_k^{(n)}\left(\frac{x}{z},\mu\right)\frac{\dd \hat \sigma_k(z,\mu)}{\dd z}=0 \, . \label{eq:Dhterms}
\end{align}
At NLO accuracy, it is enough to replace the Born contribution in the last terms in eq.~\eqref{eq:Dhterms}. Since the heavy-flavour production happens via gluon splitting, one can use the symmetry of $D_h$ and $D_{\bar{h}}$ and finally obtain the matching condition for the heavy-quark fragmentation function
\begin{align}
  &D_h^{(n)}(x,\mu)=D_{\bar{h}}^{(n)}(x,\mu)=\frac{1}{2\sigma_{h\bar{h}}} \int_x^1 \frac{\dd z}{z} D_g^{(n_L)} \left(\frac{x}{z},\mu\right) \nonumber \\
  &\left[ \frac{\dd \sigma_{h\bar{h}g^{\text{id.}}}(z,m)}{\dd z} - \frac{\dd \hat \sigma_{h\bar{h}g^{\text{id.}}}(z,\mu)}{\dd z} \right] \, . \label{eq:NLOFh}
\end{align}
In the above equation $\sigma_{h\bar h}$ is the Born cross-section for the photon-induced production of a $h\bar h$ pair. An explicit result for the elementary cross-sections in~\eqref{eq:NLOFh} can be found in \ref{sec:antenna} using ingredients computed in the antenna subtraction formalism.\\

The gluon matching condition instead can be derived by studying a process with a resolved gluon at tree-level. In \cite{Cacciari_2005} the authors considered identified gluon production in association with a ``super-heavy'' quark pair. A simple matching condition due to the different running of the coupling constant was found
\begin{align}
    &\alpha_s^{(n)}(\mu) D_g^{(n)}(x,\mu)=\alpha_s^{(n_L)}(\mu) D_g^{(n_L)}(x,\mu)+\mathcal{O}(\alpha_s^3) \, , \nonumber\\
	&D_g^{(n)}(x,\mu)= D_g^{(n_L)}(x,\mu)\left(1-\frac{T_F\alpha_s}{3\pi}\log\frac{\mu^2}{m^2}+\mathcal{O}(\alpha_s^2)\right) \, . \label{gluonNLO}
\end{align}
Here $T_F$ is the standard color normalisation of the fundamental representation. The scale choice for the strong coupling in eq.~\eqref{gluonNLO} is beyond NLO accuracy.  \\

Due to the presence of an additional mass this channel is not promising for the extraction of the threshold condition for the gluon FF at NNLO. 
At NLO there are no heavy-quark (i.e. ~bottom) effects for the super-heavy-quark (i.e. ~top) production and all the mass effects are captured by studying the gluon propagator (Fig.~2 of Ref.~\cite{Cacciari_2005}).  At NNLO it would require for example the double-virtual correction to $t\bar t j$ production or the double-real correction $t\bar t b\bar b g$, with full top-mass effects and leading bottom-mass dependence.
We focused therefore on a different process: Higgs-induced hadron production in electron-position annihilation, $e^+e^-\rightarrow h^0 \rightarrow H+X$, from which eq. ~\eqref{gluonNLO} can be re-derived. 
This process is proportional to the electron Yukawa coupling $y_e^2$ and to the Wilson coefficient $C_{h^0gg}$ for the Higgs-gluon coupling.  The hadronic cross sections take the same form as eq. ~\eqref{decNLO} and \eqref{MSNLO}, replacing the partonic cross sections which corresponding ones extracted from $h^0\to gg$ decay.  In eq.~\eqref{MSNLO} the term containing $D_h^{(n)}$  is beyond accuracy, due to the NLO matching conditions and since the production of heavy-flavour quark $h$ via Yukawa electron fusion starts at $\mathcal{O}(\alpha_s)$. Analogously, since the leading terms for the light-quark production, $\dd \sigma_{gi^{\text{id.}}\bar i}$ and $\dd \hat \sigma_{gi^{\text{id.}}\bar i}$, start at $\mathcal{O}(\alpha_s)$ and the difference $D_i^{(n)}-D_i^{(n_L)}$ is  $\mathcal{O}(\alpha_s^2)$, the terms with an identified light-quark vanish in the difference. We remain only with contributions proportional to the gluon FF at NLO. We isolate the difference between the partonic cross-sections $\dd \sigma_{g}$ and $\dd \hat \sigma_{g}$
\begin{align}
    \frac{\dd \sigma_g}{\dd z}=\frac{\dd \hat \sigma_g}{\dd z} &- \left( \frac{\dd \hat \sigma_{g,f} }{\dd z} + \frac{\dd \hat \sigma_{g^{\text{id.}}h\bar h}}{\dd z } + \left.\frac{\dd \hat \sigma_{g}}{\dd z}\right|_{\mathcal{O}(\alpha_s^0)}\right)\nonumber \\
    &+\left( \frac{\dd \sigma_{g,f} }{\dd z} + \left.\frac{\dd \sigma_{g}}{\dd z}\right|_{\mathcal{O}(\alpha_s^0)}\right)+\mathcal{O}(\alpha_s^2) \,  ,
\end{align}
where $\dd \hat \sigma_{g,f}$ and $\dd \sigma_{g,f}$ represent the virtual contributions from the massless and massive heavy-flavour loops respectively. These corrections at NLO are simple since the quark-loop can affect only the gluon self-energy as discussed in \cite{Cacciari_2005}. The difference between the Born cross-sections is encoded in the decoupling relation of the Wilson coefficient e.g.~\cite{Grozin:2011nk}
\begin{align}
  \delta_{C_{h^0gg}}^{}&\coloneqq \left.\frac{\dd \hat \sigma_{g}}{\dd z}\right|_{\mathcal{O}(\alpha_s^0)}-\left.\frac{\dd \sigma_{g}}{\dd z}\right|_{\mathcal{O}(\alpha_s^0)} \nonumber \\
  & = 2 \frac{T_F\alpha_s}{3\pi}\log \frac{\mu^2}{m^2} \sigma_0^{} \delta(1-z) \, .
\end{align}
The factor of $2$ is present since the Wilson coefficient is proportional to two powers of $\alpha_s$. The choice of the scheme for the Born cross-section $\sigma_0$ is beyond the accuracy. We can now combine all the remaining ingredients in the gluon master formula at NLO
\begin{align}
    D_g^{(n)}(x,\mu)&=D_g^{(n_L)}(x,\mu)+\frac{1}{\sigma_0}\int_x^1 \frac{\dd z}{z} D_g^{(n_L)}\left(\frac{x}{z},\mu\right) \nonumber \\
    & \times \left( \frac{\dd  \sigma_{g,f} }{\dd z} - \frac{\dd \hat \sigma_{g,f} }{\dd z} + \frac{\dd  \sigma_{g^{\text{id.}}h\bar h}}{\dd z } - \frac{\dd \hat \sigma_{g^{\text{id.}}h\bar h}}{\dd z } -\delta^{}_{C_{h^0gg}} \right)  \nonumber \\
    &=D_g^{(n_L)}(x,\mu)\left(1-\frac{T_F\alpha_s}{3\pi}\log\frac{\mu^2}{m^2}+\mathcal{O}(\alpha_s^2)\right) \, . \label{gluonNLOh} 
\end{align}
The massive virtual correction ($\dd  \sigma_{g,f}$) is zero in the decoupling scheme, while the massless one ($\dd \hat \sigma_{g,f}$) contains a pole that cancels the divergence of the corresponding real correction ($\dd \hat \sigma_{g^{\text{id.}}h\bar h} $). By applying for example the procedure described in \ref{sec:antenna} to the gluon-gluon antenna $G_3^0$, one can prove that the real corrections in the two schemes exhibit the same $z$ dependence that cancels in the difference. Therefore the combination of all the elementary cross-sections gives only a logarithmic term which cancels exactly one of the two $T_F$ contributions coming from $\delta^{}_{C_{h^0gg}}$.  Eq. ~\eqref{gluonNLOh} is in agreement with eq.~\eqref{gluonNLO}, as a consequence of the universality of the matching conditions, which do not depend on the process from which they are extracted.

\section{Master formulae at NNLO}
Following the idea of the NLO derivation described above, we consider the production of an identified hadron $H$ in $e^+e^-$ annihilation and we expand its cross-section to $\mathcal{O}(\alpha_s^2)$.
\label{sec:masterformulae}
\subsection{Light-flavour matching equation}
In the decoupling scheme, the hadron has different ways to be produced via the fragmentation of a light parton. In addition to the contributions involving only the light quarks and gluons, we should also consider the production of the fragmenting light parton in association with a heavy quark pair
\begin{align}
	\frac{\dd \sigma_H}{\dd x}=&\int_x^1 \frac{\dd z}{z} \,\bigg\{ \sum_{j\in \mathbb{I}_{n_L}} D_j^{(n_L)} (\tfrac{x}{z},\mu) \frac{\dd \sigma_j(z,\mu)}{\dd z} \nonumber \\
	&\left.+D_g^{(n_L)}(\tfrac{x}{z},\mu)\frac{\dd \sigma_{h\bar h g^{\text{id.}}}(z,\mu)}{\dd z} \right. \nonumber \\
& +\sum_{i\in  \mathbb{I}_{n_L}\smallsetminus\{g\} } D_i^{(n_L)}(\tfrac{x}{z},\mu)\frac{\dd \sigma_{h\bar h i^{\text{id.}} \bar i}(z,m)}{\dd z}\bigg\}\, . \label{dec2}
\end{align}
Here the elementary cross sections are computed considering a gluon or a light (anti-)quark as resolved parton. Eq.~\eqref{dec2} defines the prediction in the decoupling scheme, as opposed to the full massless cross-section which corresponds to the expansion of \eqref{MSNLO} up to NNLO. We recall that we indicate cross-sections in the decoupling scheme with $\dd \sigma$ while the ones in the full massless  $\overline{\text{MS}}$ scheme are denoted with $\dd \hat \sigma$. For the sake of simplicity, we will henceforth imply the dependencies of the elementary cross-sections on the variables $z$, $\mu$, $Q$ and $m$.

Following the NLO derivation, we require the difference between the two schemes to vanish. We first consider the difference of the terms
\begin{align}
    \Delta_{D_i}\coloneqq \int_x^1 \frac{\dd z}{z} \left\{ D_i^{(n_L)}(\tfrac{x}{z},\mu) \frac{\dd \sigma_i}{\dd z} -D_i^{(n)}(\tfrac{x}{z},\mu) \frac{\dd \hat \sigma_i}{\dd z} \right\} \, ,\label{diffdsigmai} 
\end{align}
respectively from eqs.~\eqref{dec2} and~\eqref{MSNLO} and focusing on a specific light quark $i\in \mathbb{I}_{n_L}\smallsetminus\{g\}$.

\begin{figure}
\begin{center}
\includegraphics[width=0.48\textwidth]{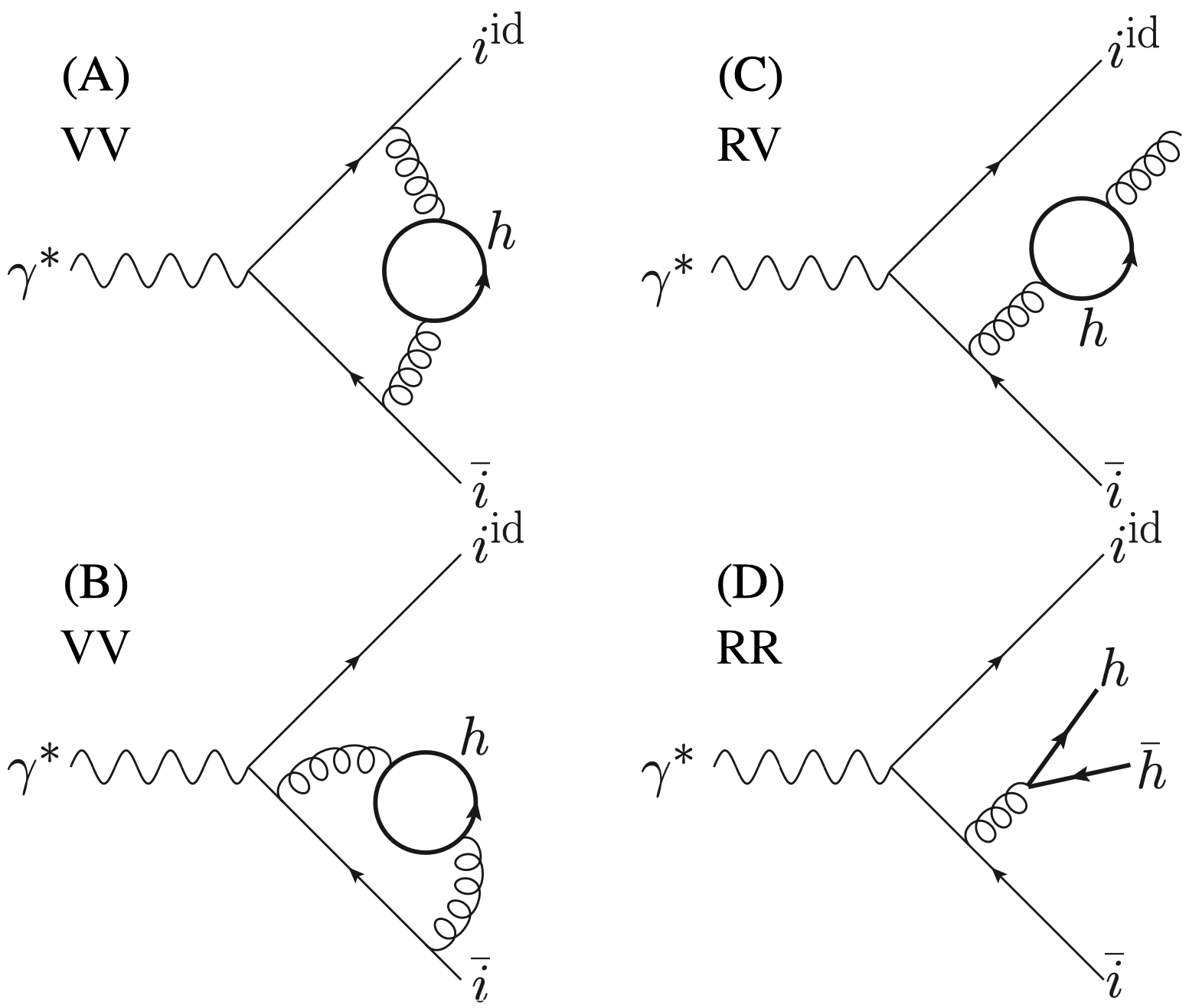}
\caption{Representative diagrams which are treated in a different way in the two schemes. The quark $h$ is massless in $\dd \hat \sigma_{i,f}$ and $\dd \hat \sigma_{h\bar h i^{\text{id.}} \bar i}$ while it has a mass $m$ in $\dd \sigma_{i,f_h}$ and $\dd \sigma_{h\bar h i^{\text{id.}} \bar i}$.} \label{fig:diagrams}
\end{center}
\end{figure}

The elementary cross-sections in~\eqref{diffdsigmai} for the production of a light quark $i$ start to differ in the two schemes at NNLO because of the presence of quark loops and real corrections with unresolved quark pairs. All pure NNLO gluonic corrections and fermionic contributions from closed loops of $n_L$ light quarks are treated in the same way.
Thus we can rewrite the massive cross-section starting from the massless one by isolating the only different diagrammatic contributions. Firstly, we subtract the NLO prediction $\dd \hat \sigma_{i}|_{\mathcal{O}(\alpha_s)}$ and all the heavy-flavour fermion terms in the massless scheme. Secondly, we add the missing heavy-quark contributions with mass dependence and the massive NLO cross-section $\dd \sigma_{i}|_{\mathcal{O}(\alpha_s)}$.
This observation is encoded in the following formula
\begin{align}
	&\frac{\dd \sigma_{i}^{}}{\dd z}=\frac{\dd \hat \sigma_i^{}}{\dd z}-\left(\frac{\dd \hat \sigma_{i,f}}{\dd z}+\frac{\dd \hat \sigma_{h\bar h i^{\text{id.}} \bar i}}{\dd z}+\left.\frac{\dd \hat \sigma_{i}^{}}{\dd z}\right|_{\mathcal{O}(\alpha_s)} \right)\nonumber \\
& +\left(\frac{\dd \sigma_{i,f}}{\dd z}+\left.\frac{\dd \sigma_{i}^{}}{\dd z}\right|_{\mathcal{O}(\alpha_s)}\right)+\mathcal{O}(\alpha_s^3)\, . \label{eq:nnlodiff}
\end{align}
Here $\dd \hat \sigma_{h\bar h i^{\text{id.}}\bar{i}}$ is the double real correction in the scheme where the quark $h$ is massless (nominally heavy). We introduce the notation $\dd \sigma_{i,f}$ for massive fermion $h$ loop corrections, given by double virtual (VV) and real-virtual (RV) corrections with some representative diagrams in Fig.~\ref{fig:diagrams}. The massless counterpart is $\dd \hat \sigma_{i,f}$ which includes all closed fermion loop contributions of a massless quark with flavour $h$.
Despite the NLO contributions being described by the same diagrams in both schemes, the strong couplings have a different running, which produces a NNLO effect. The difference reads
\begin{align}
    \delta_{\alpha_s}&\coloneqq\left.\frac{\dd \hat \sigma_{i}^{}}{\dd z}\right|_{\mathcal{O}(\alpha_s)}-\left.\frac{\dd \sigma_{i}^{}}{\dd z}\right|_{\mathcal{O}(\alpha_s)}=\left( \alpha_s^{(n)}-\alpha_s^{(n_L)} \right) \left[\frac{\dd \sigma_{i}}{\dd z}\right]^{(1)} \, ,  \label{giidiff}
\end{align}
where $[\dd \sigma_{i}]^{(1)}$ is the first-order coefficient of $\alpha_s$ in the cross-section for the production of the quark $i$. \\
We observe that the double real correction $\dd \hat \sigma_{h\bar h i^{\text{id.}}\bar{i}}$ can be decomposed in
\begin{align}
	\frac{\dd \hat \sigma_{h\bar h i^{\text{id.}} \bar i}}{dz}=\frac{\dd \hat \sigma_{h\bar h i^{\text{id.}} \bar i}^{Q_i}}{\dd z}+ \frac{\dd \hat \sigma_{h\bar h i^{\text{id.}} \bar i}^{Q_h}}{\dd z} \, ,
\end{align}
where $\dd \hat \sigma_{h\bar h i^{\text{id.}} \bar i}^{Q_i}$ is the contribution to the differential cross section proportional to $Q_i^2$ and analogously for the $Q_h^2$ term. The absence of the interference term $Q_iQ_h$ is ensured by Furry's theorem\footnote{We notice that the charge conjugation symmetry guarantees the derivation of independent matching equations for light and heavy quarks.}. An identical decomposition holds for $\dd \sigma_{h\bar h i^{\text{id.}} \bar i}$. We can separate in the difference between the two schemes the pieces proportional to $Q_i$ and $Q_h$ respectively and ask for their independent vanishing behavior.  \\

Having at hand the relation between the elementary cross-sections, we can use~\eqref{eq:nnlodiff} in~\eqref{diffdsigmai} and isolate all the contributions proportional to a specific light-quark charge. The difference between the two schemes proportional to $Q_i^2$ gives the following equation
\begin{align}
	&\int_x^1 \frac{\dd z}{z} \left\{ \left(D_i^{(n_L)} (\tfrac{x}{z},\mu) -D_i^{(n)} (\tfrac{x}{z},\mu)  \right) \frac{\dd \hat \sigma_i^{}}{\dd z} +\right. \nonumber \\
	&\left.D_i^{(n_L)} (\tfrac{x}{z},\mu)\left( \frac{\dd \sigma^{Q_i}_{h\bar h i^{\text{id.}} \bar i}}{\dd z} - \frac{\dd \hat \sigma^{Q_i}_{h\bar h i^{\text{id.}} \bar i}}{\dd z} \right)+    \right. \nonumber \\
 & \left. D_i^{(n_L)} (\tfrac{x}{z},\mu)\left( \frac{\dd \sigma_{i,f}^{}}{\dd z}-\frac{\dd \hat \sigma_{i,f}^{}}{\dd z} + \delta_{\alpha_s} \right) + \right. \nonumber \\ 
    & \left. D_g^{(n_L)}(\tfrac{x}{z},\mu)\frac{\dd \sigma_{i\bar ig^{\text{id}}}}{\dd z} - D_g^{(n)}(\tfrac{x}{z},\mu)\frac{\dd \hat \sigma_{i\bar i g^{\text{id}}}}{\dd z}\right\}=0 \, ,	\label{eqintDi}
\end{align}
which is valid $\forall i \in \mathbb{I}_{n_L}-\{g\}$.
The two terms in the last line cancel exactly at NNLO by mean of eqs.~\eqref{gluonNLO} and \eqref{giidiff}. Therefore the gluon fragmentation functions will affect the light quark matching condition only starting at $\text{N}^3\text{LO}$. In other words, at NNLO the fragmentation function $D_i^{(n)}$ can be simply written by terms convoluted with $D_i^{(n_L)}$. Since the difference between the light-quark fragmentation functions is $\mathcal{O}(\alpha_s^2)$, in the first line of eq.~\eqref{eqintDi} we only need the Born cross-section\footnote{We stress that $\sigma_{i\bar{i}}$ is the Born cross-section for the photon-induced production of a $i\bar{i}$ pair which does not depend on the mass of the quarks.}
\begin{align}
	 \frac{\dd \hat \sigma_i^{}}{\dd z} = \sigma_{i\bar{i}}\delta(1-z) +\mathcal{O}(\alpha_s) \, .
\end{align}
It follows immediately that
\begin{align}
	&D_i^{(n)} (x,\mu) = D_i^{(n_L)} (x,\mu) + \frac{1}{\sigma_{i\bar{i}}} \int_x^1 \frac{\dd z}{z} D_i^{(n_L)} (\tfrac{x}{z},\mu) \delta_{D_i}^i(z) \, , \label{DiMasterFormiula}
\end{align}
with
\begin{align}\label{eq:deltaiDi}
\delta_{D_i}^i(z)\coloneqq \frac{\dd \sigma^{Q_i}_{h\bar h i^{\text{id.}} \bar i}}{\dd z} - \frac{\dd \hat \sigma^{Q_i}_{h\bar h i^{\text{id.}} \bar i}}{\dd z} + \frac{\dd \sigma_{i,f}}{\dd z}-\frac{\dd \hat \sigma_{i,f}}{\dd z}  + \delta_{\alpha_s} \, ,
\end{align}
where the massive contributions are indicated with $\dd \sigma$ while the full massless $\overline{\text{MS}}$ counterparts with $\dd \hat \sigma$. As done for the NLO matching \cite{Cacciari_2005}, the massive differential cross-sections are computed in the small-mass limit, following the approach used for PDFs in Ref. \cite{Buza:1996wv}. Here power corrections are neglected to ensure mass factorisation in the zero-mass VFNS. In the space-like case, alternative approaches have been investigated (see for instance Ref. \cite{Bertone:2017djs} and the references therein), starting with the ACOT proposal \cite{Aivazis:1993pi} which can accomodate power corrections. Also in the time-like case power corrections can be large at $Q\sim m$ and alternative approaches in the spirit of a general-mass VFNS can be performed. \\

In Section \ref{sec:xsections} we  explicitly compute the differential cross-sections in eq.~\eqref{eq:deltaiDi} neglecting mass power-corrections and express the result in Mellin space. For this purpose we introduce the Mellin transform
\begin{align}
    \mathcal{M}_{(N,x)}\left[ f(x) \right]\coloneqq \int_0^1 \dd x \,\,x^{N-1} \,f(x) \, .
\end{align}
The matching formula \eqref{DiMasterFormiula} in Mellin space reads
\begin{align}
   D_i^{(n)}(N,\mu)=\left\{1+\frac{1}{\sigma_{i\bar{i}}}\mathcal{M}_{(N,z)}\left[\delta_{D_i}^i(z)\right]\right\}D_i^{(n_L)}(N,\mu) \, ,
\end{align}
with the Mellin transform of a fragmentation function
\begin{align}
	D_i^{(n)}(N,\mu)=\mathcal{M}_{(N,x)}\left[ D_i^{(n)}(x,\mu) \right] \, .
\end{align}
\subsection{Heavy-flavour matching equation}
\label{subsec:heavyMF}
By means of the same process considered in the derivation of the $D_i$ matching equation, we can also extract the master formula for the heavy-flavour matching. This equation can be derived requiring the vanishing of the terms proportional to $Q_h^2$ in the difference between the two predictions: the full massless scheme of eq.~\eqref{MSNLO} against the decoupling scheme of eq.~\eqref{dec2}. Focusing on the gluonic contributions 
\begin{align}
	\Delta_{D_g}^{h}\coloneqq \int_x^1 \frac{\dd z}{z} \left\{ D_g^{(n_L)}(\tfrac{x}{z},\mu)  \frac{\dd \sigma_g^{}}{\dd z}-  D_g^{(n)}(\tfrac{x}{y},\mu)  \frac{\dd \hat \sigma_g^{}}{\dd z}\right\} \, ,
\end{align}
we can use the NLO result of eq.~\eqref{gluonNLO} in order to express $D_g^{(n)}$ in the $n_L$-scheme, since the elementary cross-sections start at $\mathcal{O}(\alpha_s)$. By considering the different schemes for the couplings, we obtain
\begin{align}
	\Delta_{D_g}^{h}=\int_x^1 \frac{\dd z}{z} D_g^{(n_L)} (\tfrac{x}{z},\mu) \frac{\dd \hat \sigma_{h\bar{h}g^{\text{id.}}}}{\dd z} \, .\label{eq:DeltahDg}
\end{align}
We stress that the difference between $\dd \hat \sigma^{}_g$ and $\dd \sigma^{}_g$ does not involve quark loops in the virtual corrections at $\mathcal{O}(\alpha_s)$ and in the real correction we cannot have an additional quark pair since a gluon has to be resolved. Thus, the difference of the two elementary cross-sections at $\mathcal{O}(\alpha_s)$ is $\dd \hat \sigma_{h\bar{h}g^{\text{id.}}}$.\\
The term in eq.~\eqref{eq:DeltahDg} is combined with the massive counterpart in eq.~\eqref{dec2}.
By including all relevant contributions we obtain
\begin{align}
	\int_x^1 \frac{\dd z}{z}& \left\{ \left(D_h^{(n)}(\tfrac{x}{z},\mu)+D_{\bar{h}}^{(n)}(\tfrac{x}{z},\mu) \right)\frac{\dd \hat \sigma_{h}}{\dd z} \right. +\nonumber \\
 & D_g^{(n_L)}(\tfrac{x}{z},\mu) \left( \frac{\dd \sigma_{h\bar{h}g^{\text{id.}}}}{\dd z} - \frac{\dd \hat \sigma_{h\bar{h}g^{\text{id.}}}}{\dd z}\right)+ \nonumber \\
	&\left. \sum_{i\in \mathbb{I}_{n_L}\smallsetminus\{g\}}  D_i^{(n_L)}(\tfrac{x}{z},\mu) \left( \frac{\dd \sigma^{Q_h}_{h\bar{h}i^{\text{id.}}\bar{i}}}{\dd z} - \frac{\dd \hat \sigma^{Q_h}_{h\bar{h}i^{\text{id.}}\bar{i}}}{\dd z}\right)  \right\}=0 \, . 	\label{Dhdiff}
\end{align}
The first term above comes only from the massless prediction where the nominally heavy-flavour $h$ is a parton with its own fragmentation function. 
The terms with a resolved gluon ($\dd \sigma_{h\bar{h}g^{\text{id.}}}$ and $\dd \hat \sigma_{h\bar{h}g^{\text{id.}}}$) contribute only to the heavy quark matching condition, since there are no unresolved light quark pairs accompanied by $Q_i^2$ terms. The massless cross-section $\dd \hat \sigma^{Q_h}_{h\bar{h}i^{\text{id.}}\bar{i}}$ comes from the residual $Q_h^2$ piece in eq.~\eqref{eq:nnlodiff}. \\

To compute the matching condition for the heavy-quark fragmentation function, it is necessary to isolate the $D_h^{(n)}$ term in \eqref{Dhdiff}. This is not trivial in direct space but it can be done in Mellin space, where convolutions become simple products:
\begin{align}
	D_h^{(n)}&(N,\mu)=\left(\mathcal{M}_{(N,z)}\left[ \frac{\dd \hat \sigma_{h}}{\dd z} \right]\right)^{-1} \nonumber \\
 &\left\{ D_{g}^{(n_L)}(N,\mu)\,\, \mathcal{M}_{(N,z)} \left[  \frac{\dd \sigma_{h\bar{h}g^{\text{id.}}}}{\dd z} - \frac{\dd \hat \sigma_{h\bar{h}g^{\text{id.}}}}{\dd z}\ \right] +\right. \nonumber \\
	&\left. \sum_{i\in \mathbb{I}_{n_L}-g} D_i^{(n_L)}(N,\mu)\,\, \mathcal{M}_{(N,z)}\left[ \frac{\dd \sigma^{Q_h}_{h\bar{h}i^{\text{id.}}\bar{i}}}{\dd z} - \frac{\dd \hat \sigma^{Q_h}_{h\bar{h}i^{\text{id.}}\bar{i}}}{\dd z} \right] \right\}\, .	\label{master:Dh}
\end{align}

The massless ingredients in the above are available from \cite{Gehrmann:2022pzd, Bonino:2024adk}, while the massive ones correspond to all radiative corrections to the process $\gamma^*\to h\bar{h}$ (proportional to $Q_h$). They require integration in $d=4-2\epsilon$ dimensions of the massive squared matrix elements of \cite{Bernreuther:2013uma,Bernreuther:2011jt} with fragmentation kinematics. 
We leave the calculation of the elementary cross-sections in eq.~\eqref{master:Dh} to subsequent work. 

\subsection{Gluon matching equation}
\label{subsec:gluonMF}
As already mentioned, the derivation of the matching condition for $D_g$ requires a different process since the gluon production is suppressed by a power of $\alpha_s$ in $e^+e^-\to\gamma^*\to H+X$.  Again we compare eq. ~\eqref{decNLO} and \eqref{MSNLO} with partonic cross sections computed in $h^0\to gg$ decay with effective Higgs-gluon coupling.  We first notice that 
\begin{align}
    \Delta^g_{D_i}\coloneqq \int_x^1 \frac{\dd z}{z}  \left(D_i^{(n_L)}\frac{\dd \hat \sigma_i}{\dd z}-D_i^{(n)}\frac{\dd \sigma_i}{\dd z}\right) \, ,
\end{align}
is beyond the accuracy, indeed the elementary cross-sections starts at $\mathcal{O}(\alpha_s)$ while the difference between the fragmentation functions is $\mathcal{O}(\alpha_s^2)$. Therefore the NNLO matching condition for $D_g^{(n)}$ will be a function only of $D_g^{(n_L)}$ since the dependence on $D_i^{(n_L)}$ for $i\in \mathbb{I}_{n_L}\smallsetminus\{g\}$ is vanishing. \\

The cross-section $\dd \hat \sigma_h$ in the massless scheme can be simply computed at leading order $\mathcal{O}(\alpha_s)$, namely the Born contribution $\dd \hat \sigma_{g h^{\text{id.}}\bar h}$, since $D_h$ starts with an additional strong coupling compared to the gluon FF.  In order to isolate the gluon FF, we have to relate the gluon partonic cross-sections in the two schemes
\begin{align}
    \frac{\dd \sigma_g}{\dd z}=\,&\frac{\dd \hat \sigma_g}{\dd z} - \bigg[ \frac{\dd \hat \sigma_{g,f} }{\dd z}+ \frac{\dd \hat \sigma_{g^{\text{id.}}h\bar h}}{\dd z } -\frac{\dd \hat \sigma_{g^{\text{id.}}h\bar h,f} }{\dd z}  \nonumber \\
    &+ \left. \left(\frac{\dd \hat \sigma_{g}}{\dd z}-\frac{\dd \hat \sigma_{g^{\text{id.}}h\bar h}}{\dd z}-\frac{\dd \hat \sigma_{g,f}}{\dd z}\right)\right|_{\mathcal{O}(\alpha_s)}\bigg] \nonumber \\
    &+\left[ \frac{\dd \sigma_{g,f} }{\dd z} + \left.\left(\frac{\dd \sigma_{g}}{\dd z}-\frac{\dd \sigma_{g,f}}{\dd z}\right)\right|_{\mathcal{O}(\alpha_s)}\right]+\mathcal{O}(\alpha_s^3) \, .
\end{align}
Here $\dd \hat \sigma_{g,f}$ and its massive counterpart $\dd \sigma_{g,f}$ include all the fermionic one-loop and two-loop effects. The cross-section $\dd \hat \sigma_{g^{\text{id.}}h\bar h}$ must be computed at NLO, by including the real correction $\dd \hat \sigma_{g^{\text{id.}}gh\bar h}$. In order to avoid double counting problems between $\dd \hat \sigma_{g,f}$ and $\dd \hat \sigma_{g^{\text{id.}}h\bar h}$ we subtract the cross-section $\dd \hat \sigma_{g^{\text{id.}}h\bar h,f}$ for the heavy-flavour one-loop effects in $gh\bar h$ production, present in both contributions. As usual,  for the diagrams the are shared among the two schemes, we have to take into account the difference in the couplings
\begin{align}
    \delta_{C_{h^0gg},\alpha_s}&\coloneqq  \left. \left(\frac{\dd \hat \sigma_{g}}{\dd z}-\frac{\dd \hat \sigma_{g^{\text{id.}}h\bar h}}{\dd z}-\frac{\dd \hat \sigma_{g,f}}{\dd z}\right)\right|_{\mathcal{O}(\alpha_s)} \nonumber \\
    & -\left.\left(\frac{\dd \sigma_{g}}{\dd z}-\frac{\dd \sigma_{g,f}}{\dd z}\right)\right|_{\mathcal{O}(\alpha_s)}. \label{eq:decouplingDg}
\end{align}
For eq.~\eqref{eq:decouplingDg} we need the second-order decoupling relation for the Wilson coefficient $C_{h^0gg}$ \cite{Grozin:2011nk} and use the first-order decoupling relation for the strong coupling $\alpha_s$ in the NLO virtual and real corrections e.g.~\cite{Chetyrkin:1997un}. The difference between the two schemes reads
\begin{align}
   & \int_x^1 \frac{\dd z}{z}\left(D_g^{(n)}-D_g^{(n_L)}\right)\frac{\dd \hat \sigma_g}{\dd z} \nonumber \\
   &=\int_x^1 \frac{\dd z}{z} D_g^{(n_L)}\left\{  \frac{\dd \sigma_{g,f} }{\dd z} -\left( \frac{\dd \hat \sigma_{g,f} }{\dd z} -\frac{\dd \hat \sigma_{g^{\text{id.}}h\bar h,f} }{\dd z}\right)\right\} + \nonumber \\
    &+\int_x^1 \frac{\dd z}{z} D_g^{(n_L)} \left\{ \frac{\dd  \sigma_{g^{\text{id.}}h\bar h}}{\dd z} - \frac{\dd \hat \sigma_{g^{\text{id.}}h\bar h}}{\dd z} -\delta_{C_{h^0gg},\alpha_s} \right\} \nonumber \\
    &-\int_x^1 \frac{\dd z}{z} \left(D^{(n)}_h+D^{(n)}_{\bar h} \right)\frac{\dd \hat \sigma_{gh^{\text{id.}}\bar h} }{\dd z} \,  .
\end{align}
We can isolate the gluon FF in the massless scheme
\begin{align}
   & D_g^{(n)}(x,\mu)=D_g^{(n_L)}(x,\mu)\left(1-\frac{T_F\alpha_s^{(n_L)}}{3\pi} \log\frac{\mu^2}{m^2} \right) \nonumber \\
    &+ \frac{1}{\sigma_0}\int_x^1 \frac{\dd z}{z} D_g^{(n_L)}(\tfrac{x}{z},\mu)\left\{  \frac{\dd \sigma_{g,f} }{\dd z} -\left( \frac{\dd \hat \sigma_{g,f} }{\dd z} -\frac{\dd \hat \sigma_{g^{\text{id.}}h\bar h,f} }{\dd z}\right) \right.\nonumber \\
    &+\left. \frac{\dd  \sigma_{g^{\text{id.}}h\bar h}}{\dd z} - \frac{\dd \hat \sigma_{g^{\text{id.}}h\bar h}}{\dd z} -\delta_{C_{h^0gg},\alpha_s} +\frac{T_F\alpha_s^{(n_L)}}{3\pi} \log\frac{\mu^2}{m^2}\frac{\dd \hat \sigma_g}{\dd z} \right\}\nonumber \\
    &-\frac{1}{\sigma_0}\int_x^1 \frac{\dd z}{z} \left(D^{(n)}_h+D^{(n)}_{\bar h} \right)\frac{\dd \hat \sigma_{gh^{\text{id.}}\bar h} }{\dd z}.
\end{align}
The first line represents the NLO matching correction and the second line contains the virtual corrections due to the different treatment of massive and massless loops. The first contribution in the third line captures the differences of the two schemes in the production of the heavy-flavour pair in association to the resolved gluon, while other contributions in the same line are related to the different running of the couplings. The last term takes into account the hadron generation via heavy-flavour fragmentation which is present in the $n$-scheme and affects the gluon contribution in the fragmentation process compared to the massive scheme. Once can express the $D_g^{(n)}$ in terms of FFs in the $n_L$ scheme only, by writing $D_h^{(n)}$ via $D_g^{(n_L)}$ with eq.~\eqref{eq:NLOFh}.  \\

The massless ingredients in the above expression can in principle be extracted from gluon-gluon fragmentation antennae (the so-called $\mathcal{F}$,$\mathcal{G}$ and $\mathcal{H}$ functions) from Ref.~\cite{Bonino:2024adk}. On the other hand,  the massive loops corrections at NNLO with fragmentation kinematics and the massive NLO differential cross-section for the production of a gluon and a heavy-quark pair are currently unknown, and require a dedicated study. \\

In the rest of this paper we focus on the complete derivation of the light-quark matching condition, by computing the differential cross sections needed in eq.~\eqref{DiMasterFormiula}.

\section{Calculation of elementary cross-sections for light-quark fragmentation}
\label{sec:xsections}
\subsection{Massless contributions}
\label{sec:mless}
The massless ingredients in eq.~\eqref{eq:deltaiDi} have already been computed up to NNLO for example in the context of antenna subtraction. We introduce $\left. \mathcal{{J}}^{(2)}_{q\bar{q}}\right|_{N_f}$ as the $N_f$ part of the UV-renormalised double real and real-virtual $Q^2_i$ corrections to the production of the resolved light-quark, with mass factorisation counterterms. This contribution is obtained using some of the integrated dipoles of \cite{Bonino:2024adk} with full scale dependence as explained in \ref{sec:antenna}. The IR poles in $\left. \mathcal{{J}}^{(2)}_{q\bar{q}}\right|_{N_f}$ cancel precisely with the double virtual (VV) correction given by the $N_f$-piece of the UV-renormalised two-loop form factor $\left.F^{(2)}_q\right|_{N_f}$ of \cite{Gehrmann:2010ue}. 
We find that the massless ingredients listed in eq.~\eqref{eq:deltaiDi} are given by 
\begin{align}\label{eq:J2F2}
 \frac{\dd \hat \sigma^{Q_i}_{h\bar h i^{\text{id.}} \bar i}}{\dd z} +\frac{\dd \hat \sigma_{i,f}}{\dd z} = \sigma_{i\bar{i}}\left(\frac{\alpha_s}{2\pi}\right)^2 2 C_F\left(\left. \mathcal{{J}}^{(2)}_{q\bar{q}}\right|_{N_f}+\left.F^{(2)}_q\right|_{N_f}\right) \, .
\end{align}
\noindent Due to the normalisation conventions of the antenna functions, we multiply by a factor of $2$. The above result can be written in Mellin space using the package \texttt{MT} \cite{Hoschele:2013pvt} and in \ref{sec:Mellin} we provide its Mellin transform for $Q=\mu$ \eqref{masslessres}. This result can be checked against the perturbative corrections to the hadronic $R$-ratio, as done for instance in eq. (5.6) of Ref. \cite{Gehrmann-DeRidder:2004ttg}. We find 
\begin{align}
    \frac{1}{\sigma_{i\bar i}}\mathcal{M}_{(1,z)}\left[\frac{\dd \hat \sigma^{Q_i}_{h\bar h i^{\text{id.}} \bar i}}{\dd z} +\frac{\dd \hat \sigma_{i,f}}{\dd z}\right]&=\left(\frac{\alpha_s}{2\pi}\right)^2 C_F\left(-\frac{11}{4}+2\zeta_3\right) \, ,
\end{align}
which corresponds to the $N_f$ component of the $R$-ratio at $\mathcal{O}(\alpha_s^2)$.
Having at hand the massless ingredients we can now move to the massive ones present in eq.~\eqref{eq:deltaiDi}.

\subsection{Massive contributions}
\label{sec:msive}
In this section we explicitly compute the massive cross-section $\dd \sigma^{Q_i}_{h\bar{h}i^{\text{id.}}i}$. For this purpose we define the normalised matrix-element
\begin{align}
    \bar B_{ih\bar{h}i}^0 = \frac{|\mathcal{M}^0_{ih\bar{h}\bar{i}}|^2}{|\mathcal{M}^0_{i\bar{i}}|^2}\, ,
\end{align}
where $\mathcal{M}^0_{ih\bar{h}\bar{i}}$ and $\mathcal{M}^0_{i\bar{i}}$ are tree-level matrix elements proportional to $Q_i$ that we compute with \texttt{FeynCalc} \cite{Mertig:1990an,Shtabovenko_2020}. In \cite{Bernreuther:2011jt} the authors computed the un-integrated and the inclusively integrated contribution to $\dd \sigma^{Q_h}_{h\bar{h}i\bar{i}}$.To check our setup, we have calculated the un-integrated contribution proportional to $Q_h^2$ as well, finding agreement with \cite{Bernreuther:2011jt}. We perform the analytical integration in four space-time dimensions remaining differential in the momentum-fraction $z$ of the quark $i$. We report in \ref{sec:integration} the details of the calculation and the result for $\mathcal{\bar B}_{ih\bar{h}i}^{0,\text{id.}i}$, normalised according to \cite{Gehrmann:2022pzd}. This function is finite in four dimensions since the light-quark is resolved and the mass of the unresolved quark-pair protects it from collinear divergences. We have performed numerical checks of the analytic integration for several physical values of the parameters. In addition, as validation, we have also integrated the matrix-element using a different phase space parametrisation, as reported in \ref{sec:secondintegration}.

The differential cross-section thus reads
\begin{align}
    \frac{\dd \sigma_{h\bar{h}i^{\text{id.}}i}^{Q_i}}{\dd z}=\sigma_{i\bar{i}}\left(\frac{\alpha_s}{2\pi}\right)^2 2C_F \mathcal{\bar B}_{ih\bar{h}\bar{i}}^{0,\text{id.}i}(z,\rho) \, ,
\end{align}
where the total cross-section is obtained by integrating in $z$ from $0$ to $1-\rho$ with $\rho=\frac{4m^2}{Q^2}$. In order to perform the Mellin transform, we have to regularise the end-point soft divergences by using the heavy-quark mass as infrared regulator. We isolate the divergent term $\mathcal{\bar D}(z)$ in the function $\mathcal{\bar B}_{ih\bar{h}i}^{0,\text{id.}i}(z,\rho)$ and study its action on a test function $g(z)$ in the small-mass limit
\begin{align}
	 & \int_{0}^{1-\rho} \dd z \,\mathcal{\bar D}(z) g(z) =   \nonumber \\
	&= \int_0^1 \dd z \left\{ \left[  \mathcal{\bar D}(z) \right]_+ + \delta(1-z) \int_0^{1-\rho}\dd t\, \mathcal{\bar D}(t) \right\} g(z) +\mathcal{O}(\rho) \, .
\end{align}
The plus-prescription
\begin{align}
  \int_0^1 \dd z \,\left[  \mathcal{\bar D}(z) \right]_+ g(z) = \int_0^1 \dd z\, \mathcal{\bar D}(z) (g(z)-g(1))
\end{align}
emerges by reformulating the divergent component with the introduction of a mass-dependent contribution multiplied by $\delta(1-z)$. 
We can now perform the Mellin transform of the manipulated expression
\begin{align}
\mathcal{\bar B}_{ih\bar{h}\bar{i}}^{0,\text{id.}i}=\mathcal{\bar{B}}_{ih\bar{h}\bar{i},\text{reg.}}^{0,\text{id.}i}&+\left[  \mathcal{\bar D}(z) \right]_+  \nonumber \\
&+\delta(1-z) \int_0^{1-\rho}\dd t\, \mathcal{\bar D}(t) + \mathcal{O}(\rho) \, , \label{B40decomposition}
\end{align}
where $\mathcal{\bar{B}}_{ih\bar{h}\bar{i},\text{reg.}}^{0,\text{id.}i}(z)=\mathcal{\bar B}_{ih\bar{h}\bar{i}}^{0,\text{id.}i}(z)-\mathcal{\bar{D}}(z)$. The term proportional to $\delta(1-z)$ will be combined with the double virtual contribution. We perform the Mellin transform making use of the results in \cite{Bl_mlein_1999} and the following relation
\begin{align}
    &\mathcal{M}_{(N,z)}\left[\frac{1}{1-z}\text{Li}_2\left(\frac{z-1}{z}\right)\right]=S_{2,1}(N)+\frac{1}{N}S_2(N)  \nonumber \\
    &+S_1(N) \left(\frac{\pi}{6} ^2- S_2(N)\right)-\frac{1}{N^3}-\frac{\pi ^2}{6 N}-2 \zeta_3\,.
\end{align}
The harmonic sums $S_{i_1}$ and $S_{i_1,i_2}$ are defined as in \cite{Bl_mlein_1999}
\begin{align}
    S_{i_1,i_2,\dots,i_m} (N)=\sum_{n_1=1}^N \frac{1}{n_1^{i_1}} \sum_{n_2=1}^{n_1} \frac{1}{n_2^{i_2}}\dots \sum_{n_m=1}^{n_{m-1}} \frac{1}{n_m^{i_m}} \, , \label{eq:SiSij}
\end{align}
for $N\in \mathbb{N}$ and $i_1,i_2,\dots,i_m>0$.

In the decoupling scheme, the RV correction described by the diagram (C) of Fig. \ref{fig:diagrams} is vanishing since the UV divergences are subtracted at zero-momentum. The VV corrections represented by the diagrams (A) and (B) in Fig. \ref{fig:diagrams} were computed in \cite{Blumlein:2016xcy}. The time-like result is given by eq. (B.15) in the Appendix B where the authors have isolated the massive form-factor with the subtraction at zero-momentum for UV regularisation. We expand this result in the small mass limit and perform its Mellin transform. When combining the VV correction with the RR contribution we observe a cancellation of the third power of $\log \frac{m}{Q}$. This is a solid check of our double real correction that can be understood from the massification point of view \cite{Mitov:2006xs}. In the massless calculation in dimensional regularisation, this cancellation corresponds to the cancellation of the deepest pole $\epsilon^{-3}$.
The remainder term is $\delta_{\alpha_s}$ from eq.~\eqref{giidiff}. The NLO differential cross-section $[\dd \sigma_i]^{(1)}$ can be found in \cite{Nason:1993xx}.
Its Mellin transform reads
\begin{align}
\nonumber \\
&\mathcal{M}_{(N,z)}\left[\delta_{\alpha_s}\right]=\sigma_{i\bar{i}}\frac{\alpha_s^2}{6\pi^2}C_FT_F\log\frac{\mu^2}{m^2} \left[
\frac{3}{(N+1)^2}-\frac{2}{N^2}\right. \nonumber \\
&\left.-\frac{3}{2 (N+1)}+\frac{1}{N} -\frac{9}{2} +\left(\frac{1}{N+1}-\frac{1}{N}+\frac{3}{2}\right) S_1(N)+4 S_2(N)
\right. \nonumber \\
&\left. +2 S_{1,1}(N)+\log\frac{\mu^2}{Q^2} \left(2 S_1(N)+\frac{1}{N+1}-\frac{1}{N}-\frac{3}{2}\right) \right]\,.
\label{eq:deltaalphas}
\end{align}
We stress that the presence of this term comes from the different running of the strong couplings in the two schemes. If we had used the same UV renormalisation in the two schemes, we would have obtained eq.~\eqref{eq:deltaalphas} from the RV and VV contributions at zero-momentum. Finally we combine the massive contributions in Mellin space, and present the analytical result for $Q=m=\mu$ in eq.~\eqref{massiveresult} of \ref{sec:Mellin}.
\vspace{-0.5cm}
\subsection{Combination and results}
We are in the position of taking the difference between the massless and the massive calculation, neglecting mass power-corrections, and compute eq.~\eqref{eq:deltaiDi} in Mellin space. 
\begin{widetext}
\noindent Here we report the light-quark matching equation in $N$ space with full scale dependence
\begin{align}
&D_i^{(n)}(N,\mu)=\biggl\{1+ \left(\frac{\alpha_s}{2\pi}\right)^2 C_F \frac{1}{N^3 (N+1)^3} \biggl[ -\frac{2}{3} N^3 (N+1)^3 S_{1,2}(N)-\frac{2}{3} N^3 (N+1)^3 S_{2,1}(N) +\frac{1}{3} N^3 (N+1)^3 S_3(N) \nonumber \\
   &+\frac{5}{9} N^3 (N+1)^3
   S_2(N)+S_1(N) \left(\frac{2}{3} N^3 (N+1)^3 S_2(N)-\frac{28}{27} N^3 (N+1)^3\right)-\frac{4}{3} N^3 (N+1)^3
   \zeta_3 \nonumber \\
   &+\left(\frac{9307}{1296}-\frac{29}{108} \pi ^2\right) N^6+ \left(\frac{9307}{432}-\frac{29}{36} \pi ^2\right) N^5+ \left(\frac{3281}{144}-\frac{29}{36} \pi ^2\right) N^4+\left(\frac{10939}{1296}-\frac{29}{108} \pi
   ^2\right) N^3-\frac{5}{54} N^2-\frac{1}{9}N+\frac{1}{6}\nonumber \\
      &-\frac{8}{9} N^3 (N+1)^3 \log ^3 2+\frac{29}{9} N^3 (N+1)^3 \log ^2 2+\frac{1}{54} \left(12 \pi ^2-359\right) N^3 (N+1)^3 \log
   2 \nonumber \\
   &+\left(\frac{10}{9} N^3 (N+1)^3 S_1(N)-\frac{2}{3} N^3 (N+1)^3
   S_2(N)-\frac{1}{36} N \left(3 N^5+9 N^4+53 N^3+67 N^2+8 N-12\right)\right) \log \left(\frac{\mu ^2}{m^2}\right) \nonumber \\
&+\left(\frac{1}{12} N^2 (N+1)^2 \left(3 N^2+3 N+2\right)-\frac{1}{3} N^3
   (N+1)^3 S_1(N)\right) \log ^2\left(\frac{\mu ^2}{m^2}\right) \biggr] +\mathcal{O}(\alpha_s^3)\biggr\}D_i^{(n_L)}(N,\mu) \, . \label{finalresult}
\end{align}
   We perform the Mellin inverse and simplify the result in direct space
     \begin{align}
     &D_i^{(n)}(z,\mu)=\left\{ 1+\left(\frac{\alpha_s}{2\pi}\right)^2 \frac{C_F}{2} \left[ -\frac{67 z}{27}+\left(-\frac{z}{6}+\frac{1}{3 (1-z)}-\frac{1}{6}\right) \log ^2(z)+\left(-\frac{11 z}{9}+\frac{10}{9
   (1-z)}+\frac{1}{9}\right) \log (z)+\frac{11}{27}+ \right.\right. \nonumber \\
      &+\left( \frac{9307}{648} -\frac{10}{3}\zeta_3 -\frac{19\pi^2}{54}-\frac{16}{9} \log ^3(2)+\frac{58 }{9}\log ^2(2) +\left(\frac{4 \pi
   ^2}{9}-\frac{359}{27}\right) \log (2) \right)\delta(1-z)
   +\frac{56}{27} \left(\frac{1}{1-z}\right)_+ \nonumber \\
    & +\left(-\frac{2}{9}+\frac{22}{9}z -\frac{2}{3}\frac{1+z^2}{1-z}\log z-\left(\frac{1}{6}+\frac{2\pi^2}{9}\right)\delta(1-z)-\frac{20}{9}\left(\frac{1}{1-z}\right)_+  \right)\log\frac{\mu^2}{m^2} \nonumber\\
    &\left.\left.+\left(-\frac{1}{3}-\frac{1}{3}z+\frac{1}{2}\delta(1-z) +\frac{2}{3} \left(\frac{1}{1-z}\right)_+  \right)\log^2\frac{\mu^2}{m^2}  \right]+\mathcal{O}(\alpha_s^3)\right\} D_i^{(n_L)}(z,\mu)\, .  \label{eq:zspacefin}
\end{align}
An important validation of the calculation is the cancellation between massive and massless ingredients of all terms proportional to $\log^2 Q$ and  $\log Q$. This is expected since we are computing a universal equation which must not depend on the energy of the process from which it is extracted.
Since these equations represent the first QCD correction, the scale choice of $\alpha_s$ has only higher order effects. 
We provide the matching condition in $z$ and Mellin spaces as ancillary files to the journal submission of this paper.
\end{widetext}

\subsection{Numerical results}
\begin{figure}
    \centering
    \includegraphics[width=0.46\textwidth]{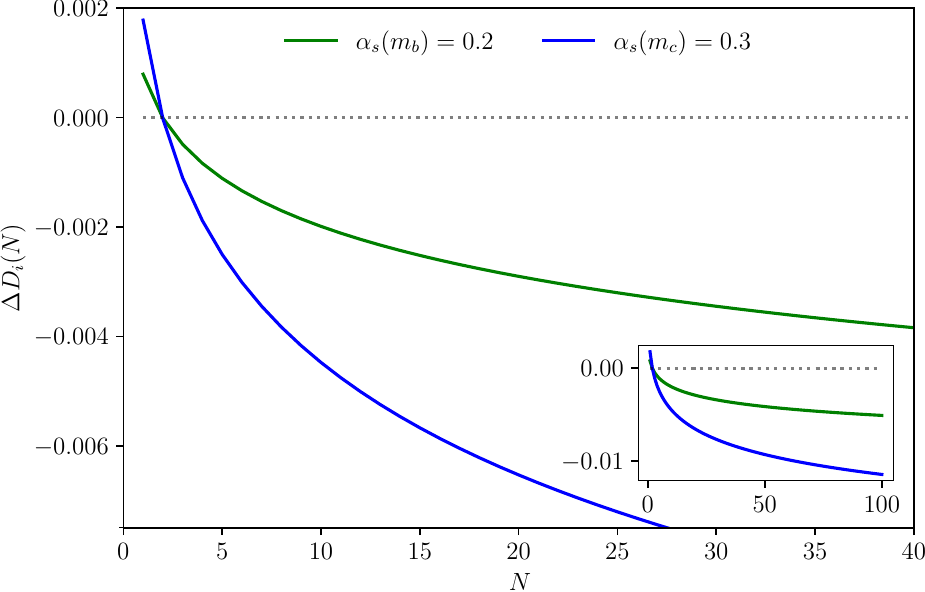}
    \caption{Light-quarks matching correction in Mellin space for values of $\alpha_s$ approximately at bottom (green) and charm (blue) thresholds.}
    \label{fig:DeltaDi}
\end{figure}

\begin{figure}
    \centering
\includegraphics[width=0.468\textwidth]{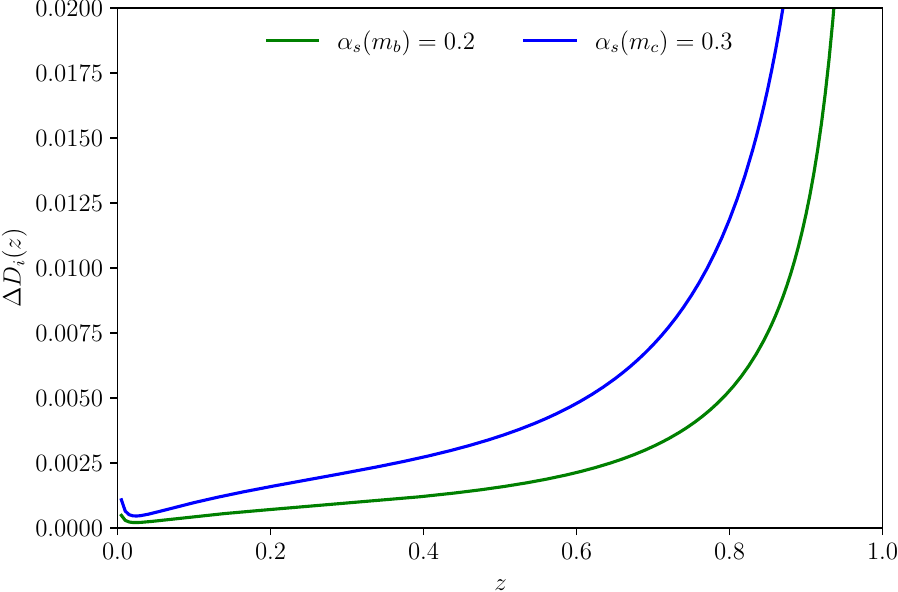}
    \caption{Light-quarks matching correction in $z$ space around bottom (green) and charm (blue) thresholds.}
    \label{fig:x_deltaDi}
\end{figure}

\noindent In order to estimate the numerical impact of the light-flavour matching condition we focus on the 
correction
\begin{equation}
    \Delta D_i(N) = \frac{D_i^{(n)}(N,m)-D_i^{(n_L)}(N,m)}{D_i^{(n_L)}(N,m)} \, ,\label{eq:correction}
\end{equation}
at the crossing scale $\mu=m$, typically set to the value of bottom or charm mass.
Due to the relative low energy scales of $\mathcal{O}(\text{GeV})$ at which the matching condition is employed, for an illustration purpose we consider the following approximated values of the strong coupling: $\alpha_s(m_b)=0.2$ and $\alpha_s(m_c)=0.3$. With this setup, the first Mellin moments of \eqref{eq:correction} are shown in Fig.~\ref{fig:DeltaDi}.
We observe a negative correction at permille level which is sensitive to the choice of the matching scale, yielding larger correction at the charm threshold (blue curve). The correction increases in absolute value with the number of the moments, driven by the double real correction $\dd \sigma^{Q_i}_{h\bar h i^{\text{id.}} \bar i}$ and its massless counterpart. Thus, the main effect in the matching of the fragmentation functions comes from a different treatment of heavy-quark radiative corrections in the two schemes.  This $N$-dependent behaviour of the matching condition cannot be given by the virtual terms which contribute only as a constant in Mellin space for $m=\mu$.  
By performing the Mellin inverse of $\Delta D_i(N)$, we provide an alternative description of the correction in the momentum fraction $z$ space as shown in Fig. \ref{fig:x_deltaDi}.


\subsection{RGE dependence}
Similarly to the space-like OMEs, the time-like matching condition satisfies a Renormalisation Group Equations (RGE). We can obtain it by taking the derivative of \begin{align}
D_i^{(n)}(N,\mu)=\left(1+\Delta D_i(N,\mu)+\mathcal{O}(\alpha_s^3)\right)D_i^{(n_L)}(N,\mu)
\end{align}
with respect to $\log\mu^2$ and use the RGEs for the fragmentation functions,
\begin{align}
    \frac{\partial}{\partial \log \mu^2} D_i^{(n_f)}=\sum_{j\in \mathbb{I}_{n_f}\smallsetminus\{g\}} \gamma_{ij}^{(n_f)} D_j^{(n_f)} + \gamma_{ig}^{(n_f)} D_g^{(n_f)} \, ,
\end{align}
for $n_f=n,n_L$. Here the anomalous couplings $\gamma_{ij}$ are the time-like splitting functions $P_{ji}$ which admit a perturbative expansion
\begin{align}
 &P_{ji}^{(n_f)}=\alpha_s^{(n_f)}P_{ii,0}^{}\delta_{ji}+\left(\alpha_s^{(n_f)}\right)^2P_{ii,1}^{(n_f)}\delta_{ji}+\mathcal{O}(\alpha_s^3) \, ,\\
 &P_{gi}=\alpha_s^{(n_f)} P_{gi,0}+\left(\alpha_s^{(n_f)}\right)^2 P_{gi,1}+\mathcal{O}(\alpha_s^3) \, .
\end{align}
The well-known tree-level and one-loop splitting coefficients can be found for example in Ref.~\cite{Ellis:1996mzs}. By expanding all the contributions in the same scheme for the coupling constant and by using the NLO gluon matching condition, the RGE for the NNLO quark matching equation reads
\begin{align}
    \frac{\partial}{\partial \log \mu^2}  \delta D_i(z,\mu,m)=\beta_0 P_{ii,0}(z) \log\frac{\mu^2}{m^2} + \hat P_{ii,1} (z)\, ,\label{eq:RGEfin}
\end{align}
where $\hat P_{ii,1}=P_{ii,1}^{(n_L)}-P_{ii,1}^{(n)}$. We verified~\eqref{eq:RGEfin} both in Mellin and direct space. Due to the Gribov-Lipatov reciprocity of the tree-level splitting coefficients, the time-like matching equation shares the same $\log^2\mu$ of the OME for the non-singlet quark PDFs~\cite{Buza:1996wv}.

\section{Conclusions and outlooks}
\label{sec:conclusions}
\vspace{-0.1cm}In this work we have studied the crossing of heavy-quark thresholds with fragmentation functions at NNLO accuracy. By considering identified hadron production in $e^+e^-$ annihilation, we derived the NNLO matching equations for the fragmentation functions of light and heavy flavours. Combining novel and existing partonic cross-sections, we computed the light-quark matching condition and provided its analytical expression in Mellin space. The present calculation should be considered as a first step towards a full NNLO description of heavy-quark effects in fragmentation functions. The calculation of the heavy-flavour matching condition, for which some analytical elementary cross-sections are still missing, is a natural extension of this work. Regarding the gluon matching condition, more conceptual efforts are needed, as its derivation requires focusing on a different process. Once the complete set of matching equations is available, a detailed assessment of their impact on fragmentation functions fits at full NNLO accuracy will be possible, shedding light on their influence on identified hadrons predictions.\\

\noindent {\bf Acknowledgements.}
We would like to thank our supervisors Thomas Gehrmann and Giulia Zanderighi for supporting our interest in this project. We are also grateful to Matteo Cacciari, Giulio Falcioni, Rhorry Gauld, Petr Jakub\v{c}\'{i}k, Matteo Marcoli, Kay Schönwald, Chiara Signorile-Signorile and Giovanni Stagnitto for fruitful discussions during the course of this work.  LB's work has received funding from the Swiss National Science Foundation (SNF) under contract 200020-204200 and from the UZH Candoc Grant scheme.

\appendix
\section{Elementary cross-sections from fragmentation antenna functions}
\label{sec:antenna}
\subsection{NLO ingredients}
The partonic cross-sections required for the NLO matching condition in eq.~\eqref{eq:NLOFh} were computed in \cite{Nason:1993xx,Furmanski:1981cw}. In this appendix we describe how to obtain the same results using integrated antenna functions. The massless contribution $\dd \hat \sigma_{h\bar{h}g^{\text{id.}}}$ can
be derived using the un-integrated $A_3^0$ antenna,
\begin{align}
A_{3}^0=\frac{|\mathcal{\hat M}^0_{hg\bar{h}}|^2}{|\mathcal{\hat M}^0_{h\bar{h}}|^2}\,,\label{eq:A30mssless}
\end{align}
with $\mathcal{\hat M}^0_{hg\bar{h}}$ tree-level matrix element for the process 
\begin{equation}
    \gamma^*\rightarrow h+\bar h+g \, ,
\end{equation} 
with all massless particles, normalised to the quark-pair production matrix element $ \mathcal{\hat M}^0_{h\bar{h}}$. The cross-section $\dd \hat \sigma_{h\bar{h}g^{\text{id.}}}$ is obtained from the fragmentation integrated antenna function $\mathcal{A}^{0,\mathrm{id.}g}_3$ defined as \cite{Gehrmann:2022pzd}
\begin{align}
    &\mathcal{A}_{hg\bar{h}}^{0,\text{id.}g}(z)=\frac{1}{C(\epsilon)}\int \dd \Phi_2^{} \frac{Q^2}{2\pi} z^{1-2\epsilon} A_{3}^0\,, \nonumber  \\
    &C(\epsilon)=\frac{\left(4\pi e^{-\gamma_E}\right)^\epsilon}{8\pi^2} , \label{def:X30int}
\end{align}
with one massless resolved gluon in $d=4-2\epsilon$ space-time dimensions. Here $\dd \Phi_2$ is the two particle phase-space for the production of the massless $h\bar{h}$ pair, $Q^2$ is the invariant mass of the process and $z$ the energy fraction carried by the gluon. We obtain the cross-section as
\begin{align}
\frac{\dd \hat{\sigma}_{h\bar{h}g^{\text{id.}}}(z,\mu)}{\dd z}={\sigma}_{h\bar{h}} \frac{\alpha_s}{2\pi}2C_F \left[ \mu^{2\epsilon} \mathcal{A}_3^{0,\text{id.}g}(z) \right]_{\text{IR-sub}} \,  \label{eq:hhgmassless} ,
\end{align}
where subtracting the infrared (IR) poles and taking the finite remainder. We stress that $\sigma_{h\bar h}$ is the Born cross-section for the electromagnetic production of a $h\bar h$ pair. We find agreement with eq.~(14) of \cite{Cacciari_2005}.

The massive counterpart $\dd \sigma_{h\bar{h}g^{\text{id.}}}$ can be obtained by considering the un-integrated massive tree-level three-parton antenna function computed in \cite{Gehrmann-DeRidder:2009lyc}
\begin{align}
A_{hg\bar{h}}^0=\frac{|\mathcal{M}^0_{hg\bar{h}}|^2}{|\mathcal{M}^0_{h\bar{h}}|^2}\,,\label{eq:A30massive}
\end{align}
with $\mathcal{M}^0_{hg\bar{h}}$ tree-level matrix element for the process \eqref{eq:process}, normalised to the quark-pair production matrix element $\mathcal{M}^0_{h\bar{h}}$. 
To obtain $\dd \sigma_{h\bar{h}g^{\text{id.}}}$ of eq.~\eqref{decNLO} we define the massive fragmentation antenna function $\mathcal{A}_{hg\bar{h}}^0$ as~\eqref{def:X30int} but using the un-integrated antenna \eqref{eq:A30massive} and the two massive particle phase-space $\dd \Phi_2^{(m,m)}$. We can explicitly perform this integration by using the phase space parametrisation from eq. (5.8) of \cite{Gehrmann-DeRidder:2009lyc}, but only integrating over the angular variable as done for the massless case in eq. (5.3) of \cite{Gehrmann:2022pzd}.
In such a way we re-derive the differential cross section of eq. (6) in \cite{Cacciari_2005} using the integrated fragmentation antenna function
\begin{align}
&\frac{\dd \sigma_{h\bar{h}g^{\text{id.}}}(z)}{\dd z} =\sigma_{h\bar{h}} \frac{\alpha_s}{2\pi}2 C_F \mathcal{A}_{hg\bar{h}}^{0,\text{id.}g}(z)=\sigma_{h\bar{h}}\frac{\alpha_s}{2\pi}2C_F \cdot \nonumber \\
&\frac{1+(1-z)^2}{z} \left[ -\log\frac{m^2}{Q^2}+\log(1-z)-1\right] +\mathcal{O}\left(\frac{m}{Q}\right)\,. \label{eq:expldec}
\end{align}
\subsection{NNLO massless ingredients}
In \cite{Gehrmann:2022pzd} the antenna formalism was extended to identified hadron production by computing the final-final fragmentation antenna functions, needed to account for all possible final-state singularities. It is customary to collect and organise the integrated antenna functions that appear in virtual subtraction terms for explicit singularities in the so-called integrated dipoles~\cite{Currie:2013vh,Chen:2022clm,Chen:2022ktf,Gehrmann:2023dxm}. Within the dipoles, mass factorisation is performed by properly subtracting NLO and NNLO mass factorisation kernels. In \cite{Bonino:2024adk}, in which the antenna subtraction method was extended to hadron fragmentation processes in hadronic collisions up to NNLO, the final-final fragmentation antenna functions were collected in final-final fragmentation integrated dipoles. We refer the reader to \cite{Bonino:2024adk} for a more in-depth discussion.

The relevant dipoles for the calculation of $\left. \mathcal{{J}}^{(2)}_{q\bar{q}}\right|_{N_f}$ in~\eqref{eq:J2F2} are the once contributing to the $N_f$-piece of the identity-preserving quark-antiquark dipole, namely
\begin{align}
     J_2^{(1)}\left(z\right)&=\left(\frac{Q^2}{\mu_R^2}\right)^{-\epsilon}\mathcal{A}_3^{0,\,\mathrm{id.}q}(z)-\left(\frac{\mu_F^2}{\mu_R^2}\right)^{-\epsilon}\Gamma^{(1)}_{qq}(z) \, , \nonumber \\
    \hat{{J}_2}^{(2)}\left(z\right)&=\left(\frac{Q^2}{\mu_R^2}\right)^{-2\epsilon}\mathcal{B}_4^{0,\,\mathrm{id.}q}(z)+\left(\frac{Q^2}{\mu_R^2}\right)^{-2\epsilon}\mathcal{{\hat{A}}}_3^{1,\,\mathrm{id.}q}(z)  \nonumber \\
    &+\frac{b_{0,F}}{\epsilon}\left(\frac{Q^2}{\mu_R^2}\right)^{-2\epsilon}\mathcal{A}_3^{0,\,\mathrm{id.}q}(z)-\left(\frac{\mu_F^2}{\mu_R^2}\right)^{-2\epsilon}\wh{\overline{\Gamma}}^{(2)}_{qq}(z) \, . \label{eq:dipoles}
\end{align}
In the above, $\mathcal{A}_3^{0,\,\mathrm{id.}q}$ is the three-parton tree-level antenna function defined in \eqref{def:X30int} while the three-parton one-loop and four-parton tree-level antennae, $\mathcal{{\hat{A}}}_3^{1,\,\mathrm{id.}q}$ and $\mathcal{B}_4^{0,\,\mathrm{id.}q}$ respectively, are defined as
\begin{align}
    \mathcal{\hat A}_3^{1,\text{id.}q}(z)&=\frac{1}{C(\epsilon)}\int \dd \Phi_2 \frac{Q^2}{2\pi} z^{1-2\epsilon} \hat A_3^1 \, , \\
    \mathcal{B}_4^{0,\text{id.}q}(z)&=\frac{1}{[C(\epsilon)]^2}\int \dd \Phi_3 \frac{Q^2}{2\pi} z^{1-2\epsilon} B_4^0\, .   \label{eq:X40def}
\end{align}
In the above $\dd \Phi_2$ and $\dd \Phi_3$ are the massless two-particle and three-particle phase spaces. 
The antennae $\hat A_3^1$ and $B_4^0$ encode the information from the three-parton one-loop and four-parton tree-level matrix elements. Precisely, the $\mathcal{{\hat{A}}}_3^{1,\,\mathrm{id.}q}$ antenna contains the fermionic one-loop correction with a gluonic radiation, e.g. diagram (C) of Fig. \ref{fig:diagrams}, while the $\mathcal{B}_4^{0,\,\mathrm{id.}q}$ encodes all double real corrections related to the production of a secondary $q\bar q$ pair of different flavour, as shown in diagram (D) of Fig.\ref{fig:diagrams}. The one-loop $\Gamma^{(1)}_{qq}$ and two-loop $\wh{\overline{\Gamma}}^{(2)}_{qq}$ mass factorisation kernels are colour-stripped splitting kernels, defined from standard evolution kernels and given in \cite{Gehrmann:2022pzd,Bonino:2024adk}. Ultimately, $b_{0,F}=-1/3$ is the fermionic contribution to the first coefficients of the QCD $\beta$-function. The renormalisation and factorisation scales $\mu_R$ and $\mu_F$ are set to the same value $\mu$ for the purpose of this work. In~\eqref{eq:dipoles} we imply the dependence of the dipoles on the scales. We stress that all the ingredients appearing in the dipole are functions of the dimensional regulator $\epsilon$. Having performed subtraction of infrared singularities by combing double real (RR), single virtual and mass-factorisation contributions, the needed dipole is given by the sum \cite{Bonino:2024adk}
\begin{equation}
    \left. \mathcal{{J}}^{(2)}_{q\bar{q}}\right|_{N_f}=\hat{{J}_2}^{(2)}-\frac{b_{0,F}}{\epsilon} J_2^{(1)}\,.
\end{equation}

\begin{widetext}
\section{Massless and massive differential cross-sections in Mellin space}
\label{sec:Mellin}
We provide results for the massless (full $\overline{\text{MS}}$ scheme) and massive (decoupling scheme) differential cross sections without scale dependence. For $Q=\mu$, we have
\begin{align}
   \mathcal{M}_{(N,z)}&\left[\frac{\dd \hat \sigma^{Q_i}_{h\bar h i^{\text{id.}} \bar i}}{\dd z} +\frac{\dd \hat \sigma_{i,f}}{\dd z}\right]= \sigma_{i\bar{i}}\left(\frac{\alpha_s}{2\pi}\right)^2 C_F \frac{1}{N (N+1)}  \left[\frac{1}{18} \left(-29 N^2-29 N+6\right) S_{1,1}(N)-\frac{2}{3} N (N+1)
   S_{2,1}(N) \right.  \nonumber \\
   &\left. -\frac{2}{3} N (N+1) S_{1,1,1}(N)-\frac{1}{3} N (N+1) S_3(N)-\frac{8}{3} N (N+1)
   S_2(N) + \frac{1}{108 N (N+1)}\bigg(\left(12 \pi ^2-247\right) N^4 \right. \nonumber \\
   &\left. +4 \left(6 \pi ^2-119\right) N^3+\left(12 \pi
   ^2-169\right) N^2+168 N+36\bigg) S_1(N) + \frac{1}{432 N^2 (N+1)^2}\bigg(\left(1371-36 \pi ^2\right) N^6 \right. \nonumber \\
   &\left.+\left(4341-108 \pi ^2\right) N^5+\left(4409-132
   \pi ^2\right) N^4+\left(2599-84 \pi ^2\right) N^3 -8 \left(3 \pi ^2-289\right) N^2+864 N+72 \bigg) \right. \nonumber \\
   & +2 N (N+1) \zeta_3 \biggr]\,,
   \label{masslessres}
\end{align}
with harmonic sums as defined in eq.~\eqref{eq:SiSij}. The massive counterpart for $Q=\mu=m$ reads
\begin{align}
   \mathcal{M}_{(N,z)}&\left[\frac{\dd \sigma^{Q_i}_{h\bar h i^{\text{id.}} \bar i}}{\dd z} +\frac{\dd \sigma_{i,f}}{\dd z}\right]=\sigma_{i\bar{i}}\left(\frac{\alpha_s}{2\pi}\right)^2 C_F \frac{1}{N^3 (N+1)^3} \bigg[ -\frac{2}{3} N^3 (N+1)^3 S_{1,2}(N)-\frac{4}{3} N^3 (N+1)^3 S_{2,1}(N) \nonumber \\
   &-\frac{2}{3} N^3
   (N+1)^3 S_{1,1,1}(N) -\frac{1}{18} N^2 \left(29 N^2+29 N-6\right) (N+1)^2 S_{1,1}(N)-\frac{19}{9} N^3 (N+1)^3 S_2(N)  +\bigg(\frac{2}{3} N^3 (N+1)^3 S_2(N)  \nonumber \\
   & +\frac{N (N+1)}{108} \big(\left(12 \pi ^2-359\right) N^4+4 \left(6 \pi ^2-175\right) N^3+\left(12
   \pi ^2-281\right) N^2+168 N+36\big)\bigg) S_1(N) \nonumber \\
   &+\frac{1}{324} \bigg( \left(3355-114 \pi ^2\right) N^6-6 \left(57 \pi ^2-1706\right)
   N^5 +\left(10689-360 \pi ^2\right) N^4+\left(4684-150 \pi ^2\right) N^3 \nonumber  \\
   & -6 \left(3 \pi ^2-284\right) N^2+612 N+108\bigg)+\frac{2}{3} N^3 (N+1)^3
   \zeta_3-\frac{8}{9} N^3 (N+1)^3 \log ^3 2 +\frac{29}{9} N^3 (N+1)^3 \log ^2 2 \nonumber \\
   & +\frac{1}{54} \left(12 \pi ^2-359\right) N^3 (N+1)^3 \log 2\bigg]\,. \label{massiveresult}
\end{align}
To simplify the expressions in Mellin space, we have written the harmonic sums in terms of a fixed argument by using their definition in eq.~\eqref{eq:SiSij}. The contributions proportional to the digamma function $\psi^{(0)}$ and its first and second-order derivatives, respectively $\psi^{(1)}$ and $\psi^{(2)}$, are written in terms of the harmonic sums by using the following relations
\begin{align}
    &\psi^{(0)}(N)=-\gamma_E +S_1(N-1)\,,\hspace{0.3cm} \psi^{(1)}(N)=\frac{\pi^2}{6} - S_2(N-1)\,,\hspace{0.3cm} \psi^{(2)}(N)=-2\zeta_3 + 2 S_3(N-1)\,. 
\end{align}
\end{widetext}

\section{Integration of the massive \texorpdfstring{$\bar B_{ih\bar{h}\bar{i}}^0$}{Bbihhbib} antenna}
\label{sec:integration}
In order to integrate the massive ingredient $\bar B_{ih\bar{h}\bar{i}}^0$, we firstly have to derive a suitable phase-space (PS) measure. In the antenna formalism, this is the first and simplest integration of a massive antenna $X_4^0$ with a fragmentation kinematics. For generality, in the first part of the derivation we avoid fixing the space-time dimension of the phase space, since the latter can be of use in the future for the integration of other massive antennae which are divergent in four dimensions and useful for a subtraction scheme. In order to simplify the problem, we consider the crossing-symmetric scattering
\begin{align}
	q+(-k_p)^{(0)} \rightarrow k_1^{(m)} + k_2^{(m)} + k_3^{(0)} \, , \label{eq:crosssym}
\end{align}
where $k_p$ is the momentum of the massless identified quark.
We have to integrate the antenna in the full three-particle phase space of eq.~\eqref{eq:crosssym},
\begin{align}
	\dd\Phi_3=\,&\frac{\dd^d k_1}{(2\pi)^d}\frac{\dd^d k_2}{(2\pi)^d}\frac{\dd^d k_3}{(2\pi)^d} (2\pi)\delta(k_1^2-m^2)(2\pi)\delta(k_2^2-m^2) \nonumber  \\
	&(2\pi) \delta(k_3^2) (2\pi)^d \delta^{(d)}(q-k_p-k_1-k_2-k_3) \nonumber \\
	=\,& (2\pi)^{-2d+3} \dd s_{12} \dd s_{23} \dd^d k_1 \dd^d k_3 \delta(k_1^2-m^2) \delta(k_3^2)   \nonumber \\
	&\delta((q-k_p-k_1-k_2)^2-m^2) \delta(s_{12}-(k_1+k_2)^2)\nonumber\\
    &\delta(s_{23}+m^2-(k_2+k_3)^2) \, .
\end{align}
In the center of mass of $q+(-k_p)$, we can express the $\delta$-function for the on-shell condition of $k_2$ in terms of the angle $z=\cos\theta_{13}$,
\begin{align}
	\delta((q-k_p-k_1-k_2)^2-m^2)=\frac{1}{2k_3^0 \sqrt{(k_1^0)^2-m^2}}\delta\left(z-\bar{z}\right) \, ,
\end{align}
where
\begin{align}
	\bar{z}=\frac{s_{12}s_{23}+s(s_{12}+s_{23})-s^2}{(s-s_{12})\sqrt{(s-s_{23})^2-4 m^2 s}} \, ,
\end{align}
with
\begin{align}
	s\coloneqq (q-k_p)^2=Q^2(1-z)\,.
\end{align}
Using polar coordinates for the $d$-dimensional measures of $k_1$ and $k_3$, we obtain
\begin{align}
	&\dd\Phi_3=\dd s_{12} \dd s_{23} \dd \Omega_{1}^{(d-2)} \dd \Omega_{3}^{(d-1)}\frac{1}{32s} \nonumber \\
    &\left[\frac{(m^2+s_{23})(-m^2 s+s_{12}(s-s_{12}-s_{23}))}{4s}\right]^{\frac{d-4}{2}}(2\pi)^{-2d+3} \, . \label{eq:phi3interm}
\end{align}
The condition on $z=\cos\theta_{13}$ fixes the coordinate system in such a way that the remaining integration over the solid angle of particle $1$ is arbitrary and can immediately be performed. Regarding the measure $\dd \Omega_{3}^{(d-1)}$, we can parametrise it by introducing two angles, $\theta$ and $\phi$, which represent the polar and azimuthal angles of the unresolved massless particle with respect to the identified quark. It yields
\begin{align}
	\dd\Phi_3=&\,\frac{s^{1-\frac{d}{2}}}{(4\pi^2)^{d}\,\Gamma(3-d)} \dd s_{12} \dd s_{23} \nonumber \\
    &\left[(m^2+s_{23})(-m^2 s+s_{12}(s-s_{12}-s_{23}))\right]^{\frac{d-4}{2}} \nonumber\\
	&\dd \cos\theta\, (1-\cos^2\theta)^{\frac{d-4}{2}} \dd \cos\phi\, (1-\cos^2\phi)^{\frac{d-5}{2}} \, .
\end{align}
In four dimension, we simply have
\begin{align}
	\int \dd\Phi_3^{(d=4)}=\frac{1}{512 \pi^4 s}\int_{4m^2}^{s} \dd s_{12} \int_{s_{23}^-}^{s_{23}^+}\dd s_{23} \int_{-1}^1\dd\cos\theta \int_{0}^{2\pi}\dd \phi \, .
\end{align}
The limits of integration $s_{23}^\pm$ can be derived computing the energy of particle $2$ in two different ways,
\begin{align}
	\begin{cases}
		\left(k_2^0\right)^2=\left( \sqrt{s}-k_1^0 -k_3^0 \right)^2,\\
		\left(k_2^0\right)^2=\left(k_1^0\right)^2+\left(k_3^0\right)^2+2\sqrt{\left(k_1^0\right)^2+m^2} \,k_3^0\, \bar{z}\, .
	\end{cases}
\end{align}
By equalising the two right-hand sides written in terms of the invariants $s,s_{12}$ and $s_{23}$, we find
\begin{align}
	s_{23}^{\pm}=\frac{(s-s_{12})\left(s_{12}\pm \sqrt{s_{12}(s_{12}-4m^2)}\right)}{2 s_{12}}\, .
\end{align}
Thus the $\mathcal{\bar B}_{ih\bar{h}\bar{i}}^{0,\text{id.}i}(z)$ in four dimension is given by the integral
\begin{align}
	\mathcal{\bar B}_{ih\bar{h}\bar{i}}^{0,\text{id.}i}(z)=\,&\frac{z}{16 \pi (1-z)} \int_{4m^2}^{s} \dd s_{12} \int_{s_{23}^-}^{s_{23}^+}\dd s_{23} \nonumber \\ 
    &\int_{-1}^1 \dd\cos\theta \int_{0}^{2\pi}\dd \phi \,\bar B_{ih\bar{h}\bar{i}}^0 \, .	\label{antennasecondpar}
\end{align}
After performing the above integrals, we obtain the so-called fragmentation antenna in an analytical form. We check numerically the correctness of the integration considering several physical PS points.\\
\\
\begin{widetext}
\noindent In conclusion, the result is expanded in the small mass limit and we obtain 
\begin{align}
  \mathcal{\bar B}_{ih\bar{h}\bar{i}}^{0,\text{id.}i}(z)= &\,\frac{1}{216 (z-1)} \bigg[-18 \left(z^2+1\right) \log ^2\frac{m^2}{Q^2}+6 \left(-13 z^2+6 \left(z^2+1\right) \log (1-z)+6 \left(z^2+1\right) \log (z)-16\right)
   \log \frac{m^2}{Q^2} \nonumber\\
   &\left. +36 \left(z^2+1\right) \text{Li}_2\left(\frac{z-1}{z}\right)+6 \pi ^2 \left(z^2+1\right)-18 \left(z^2+1\right) \log
   ^2(1-z) +6 \log (1-z) \left(13 z^2-6 \left(z^2+1\right) \log (z)+16\right) \right. \nonumber\\
   & -z (115 z+72)+12 (z (8 z+3)+8) \log (z)-172\bigg]+\mathcal{O}\left(\frac{m}{Q}\right).	 \label{Bbar40int}
\end{align}
\end{widetext}
\noindent In the next section we provide an alternative integration of the massive double real antenna as a validation of the previous result.
\section{Alternative integration of \texorpdfstring{$\bar B_{ih\bar{h}\bar{i}}^0$}{Bbihhbib} antenna}
\label{sec:secondintegration}
Here we derive an alternative parametrisation of the PS for the four-particle production with a massive pair and a resolved massless quark. We consider the scattering process for the production of two pair of particles with masses $m_1$ and $m_2$
\begin{align}
	q \rightarrow k_1^{(m_1)} + k_2^{(m_1)}+ k_3^{(m_2)} +k_p^{(m_2)}\, .
\end{align}
We will later take the limit $m_2\rightarrow 0$ in order to obtain the needed PS. We first introduce the invariant variables
\begin{align}
	y=\frac{(k_1+k_2)^2}{Q^2},\,\,z=\frac{2 q\cdot k_p}{Q^2},\,\,x=\frac{2 q\cdot k_3}{Q^2} \, ,
\end{align} 
and we want to remain differential in the variable $z$. We work in the frame of the pair of quarks with mass $m_1$, i.e. the three-momenta satisfy the condition $\mathbf{k_1}+\mathbf{k_2}=0$. In this frame the energies of the final states are
\begin{align}
	&k_1^0=k_2^0=\sqrt{Q^2} \frac{\sqrt{y}}{2}\, ,\\
	&k_p^0=\frac{\sqrt{Q^2}}{2\sqrt{y}} (1-x-y)\, ,\\
	&k_3^0=\frac{\sqrt{Q^2}}{2\sqrt{y}} (1-z-y) \, .
\end{align}
We want to parametrise the PS measure in terms of the three invariants and the two angles $(\theta,\phi)$ that describe the direction of the quark with mass $m_1$ in its quark-pair frame. We start from

\begin{align}
	\int \dd \Phi_4=&\, \int \frac{\dd^3 k_p}{(2\pi)^3 2 k_p^0} \frac{\dd^3 k_3}{(2\pi)^3 2 k_3^0}  \frac{\dd^3 k_1}{(2\pi)^3 2 k_1^0}  \frac{\dd^3 k_2}{(2\pi)^3 2 k_2^0} \nonumber\\
    & (2\pi)^4 \delta^{(4)}(q-r-r'-p-p') \nonumber \\
	=\,&\frac{Q^4}{(4\pi)^6} \int_{\rho_1}^{(1-\sqrt{\rho_2})^2}\dd y \sqrt{1-\frac{\rho_1}{y}} \int_{\sqrt{\rho_2}}^{z_+} \dd z  \nonumber
 \\ &\int_{x_-}^{x_+} \dd x \int_0^1 \dd v \int_0^{2\pi} \dd\phi	\, ,\label{NOparam}
\end{align}
where
\begin{align}
	&v=\,\frac{1}{2}(1-\cos\theta),\,\,\rho_i=\frac{4 m_i^2}{Q^2}, \,\,z_+=1-y-\sqrt{y\rho_2},\\
	&x_\pm=\,\frac{1}{4(1-y)+\rho_2}\left[(2-y)(2+\rho_2-2y-2z) \right. \nonumber \\
     &\hspace{1cm}\left.\pm 2\sqrt{(y^2-\rho_2)[(y-1+z)^2-\rho_2 y]}\right] \, .
\end{align}

In the appendix A of \cite{Nason:1997nw}, the authors derived the PS for four particles with the same mass: this is consistent with our parametrisation in eq.~\eqref{NOparam} when $m_1=m_2=m$.\\

After setting the mass $m_2$ to zero, we need to change the order of the  $y$ and $z$ integrations since we want to be differential in $z$, i.e. the energy of the quark $k_p$ measured in the $q$-frame,
\begin{align}
	\int_{\rho_1}^{1}\dd y \sqrt{1-\frac{\rho_1}{y}} \int_{0}^{1-y} \dd z=\int_{0}^{1-\rho_1} \dd z \int_{\rho_1}^{1-z} \dd y \sqrt{1-\frac{\rho_1}{y}} \, .
\end{align}
Thus we obtain
\begin{align}
	\int \dd \Phi_4=&\,\frac{Q^4}{(2\pi)^6} \int_0^{1-\rho_1} \dd z \int_{\rho_1}^{1-z} \dd y  \sqrt{1-\frac{\rho_1}{y}}\nonumber \\
 &\, \int_{1 - z - y}^{\frac{1-z-y}{1-z}} \dd x \int_{0}^1 \dd v \int_0^{2\pi} \dd\phi \, .	\label{firstparam}
\end{align}
The integrated fragmentation antenna~\eqref{antennasecondpar} using this alternative parametrisation reads as
\begin{align}
	\mathcal{\bar B}_{ih\bar{h}\bar{i}}^{0,\text{id.}i}(z)=&\,\frac{Q^4}{8\pi} \int_{\rho_1}^{1-z} \dd y \sqrt{1-\frac{\rho_1}{y}}\int_{1-z-y}^{\frac{1-z-y}{1-z}}\dd x \nonumber  \\
 &\,\int_0^1 \dd \cos\theta \int_0^{2\pi} \dd\phi \, \bar B_4^0 \, . \label{antennafirstpar}
\end{align}

\noindent A first check is the comparison of the volumes obtained by using the two PS descriptions. In the massless case, the parametrisations of eq.~\eqref{antennasecondpar} and eq.~\eqref{antennafirstpar} both give the following volume
\begin{align}
	\mathcal{V}^{}_{\text{massless}}= \frac{Q^4}{8 }(1-z)z \, .
\end{align}
By restoring the mass dependence, we again observe that the two PS parametrisations give the same volume, namely
\begin{align}
	&\mathcal{V}^{}_{\text{massive}}=\frac{Q^4 z}{16(1-z)} \left\{\ 2m^2\left(4-4z+\frac{4m^2}{Q^2}\right) \right. \nonumber\\
	&\hspace{0.5cm}\left[2\log\left(1+\sqrt{1-\frac{4m^2}{Q^2(1-z)}}\right) +\log\left(\frac{4m^2}{Q^2(1-z)}\right)\right] \nonumber\\
 &\hspace{0.5cm}\left. +\left(2-2z+\frac{4m^2}{Q^2}\right)(1-z)\sqrt{1-\frac{4m^2}{Q^2(1-z)}}\right\} \, .
\end{align}
After checking the volume, we perform a numerical integration of the $\bar{B}_4^0$ using the second parametrisation of eq.~\eqref{antennafirstpar} and find agreement with the analytic result of eq.~\eqref{Bbar40int}. 

\printbibliography

\end{document}

%% file: definitions.tex
\def\to{\rightarrow}

